\documentclass[fleqn,usenatbib]{rasti}

\usepackage{newtxtext,newtxmath}

\usepackage[T1]{fontenc}

\DeclareRobustCommand{\VAN}[3]{#2}
\let\VANthebibliography\thebibliography
\def\thebibliography{\DeclareRobustCommand{\VAN}[3]{##3}\VANthebibliography}

\usepackage{graphicx}	
\usepackage{amsmath}	
\usepackage{newtxtext,newtxmath}
\usepackage[dvipsnames]{xcolor}
\usepackage{balance}
\usepackage{orcidlink}

\usepackage{enumitem}

\def\be{\begin{equation}}
\def\ee{\end{equation}}
\def\bea{\begin{eqnarray}}
\def\eea{\end{eqnarray}}
\def\edit{}

\hypersetup{
    colorlinks = true,
    citecolor = {MidnightBlue},
    linkcolor = {BrickRed},
    urlcolor = {BrickRed}
}

\usepackage{soul}

\title[RHINO 21cm global signal experiment]{RHINO: A large horn antenna for detecting the 21cm global signal}

\author[P. Bull et al.]{Philip Bull$^{1,2}$\thanks{E-mail: \url{phil.bull@manchester.ac.uk}}\,\orcidlink{0000-0001-5668-3101},
Ahmed El-Makadema$^{1}$\,\orcidlink{0000-0001-8718-4145},
Hugh Garsden$^{1}$\,\orcidlink{0009-0001-3949-9342},
John Edgley$^{1}$,
Neil Roddis$^{1}$,
\newauthor
Jens Chluba$^{1}$\,\orcidlink{0000-0003-3725-6096},
Christopher J. Conselice$^{1}$\,\orcidlink{0000-0003-1949-7638},
Sohini Dutta$^{1}$\,\orcidlink{0000-0003-1140-1582},
Katrine A.\ Glasscock$^{1}$\,\orcidlink{0000-0001-6894-0902},
\newauthor
Ainulnabilah Nasirudin$^{1}$\,\orcidlink{0000-0003-2213-4547},
Jordan Norris$^{1}$\,\orcidlink{0009-0009-1640-9349},
Michael J.\ Wilensky$^{3,1}$\,\orcidlink{0000-0001-7716-9312},
Isabelle Ye$^{1}$\,\orcidlink{0009-0007-1958-3364},
Zheng Zhang$^{1}$\,\orcidlink{0000-0002-9154-2803}
\\
$^{1}$Jodrell Bank Centre for Astrophysics, University of Manchester, Manchester M13 9PL, UK\\
$^{2}$Department of Physics and Astronomy, University of Western Cape, Cape Town 7535, South Africa\\
$^{3}$Department of Physics and Trottier Space Institute, McGill University, 3600 University Street, Montreal, QC H3A 2T8, Canada
}

\date{Accepted XXX. Received YYY; in original form ZZZ}

\pubyear{\the\year{}}

\begin{document}
\label{firstpage}
\pagerange{\pageref{firstpage}--\pageref{lastpage}}
\maketitle

\begin{abstract}
The sky-averaged brightness temperature of the 21cm line from neutral hydrogen provides a sensitive probe of the thermal state of the intergalactic medium, particularly before and during Cosmic Dawn and the Epoch of Reionisation. This `global signal' is faint, on the order of tens to hundreds of millikelvin, and spectrally relatively smooth, making it exceedingly difficult to disentangle from foreground radio emission and instrumental artefacts. In this paper, we introduce RHINO, an experiment based around a large horn antenna operating from $60-85$~MHz. Horn antennas are highly characterisable and provide excellent shielding from their immediate environment, which are potentially decisive advantages when it comes to the beam measurement and modelling problems that are particularly challenging for this kind of experiment. The system also includes a novel continuous wave calibration source to control correlated gain fluctuations, allowing continuous monitoring of the overall gain level without needing to rapidly switch between the sky and a calibration source. Here, we describe the basic RHINO concept, including the antenna design, EM simulations, and receiver electronics. We use a basic simulation and analysis pipeline to study the impact of the limited bandwidth on recovery of physical 21cm global signal model parameters, and discuss a basic calibration scheme that incorporates the continuous wave signal. Finally, we report on the current state of a scaled-down prototype system under construction at Jodrell Bank Observatory.
\end{abstract}

\begin{keywords}
data~methods -- instrumentation -- cosmology: reionization 
\end{keywords}



\section{Introduction} \label{sec:intro}

Neutral hydrogen is one of the few truly ubiquitous cosmic tracers. It exists from around the time of last scattering at redshift $z \approx 1090$ and persists as one of the very few luminous sources through the cosmic Dark Ages \citep{2006PhR...433..181F, 2008PhRvD..78j3511P, 2013JCAP...11..066F, 2023NatAs...7.1025M}. After that, it is ionised in a patchy, localised fashion around the first stars and galaxies during Cosmic Dawn ($z \lesssim 30$), eventually becoming completely ionised on large scales by the end of the Epoch of Reionisation ($z \lesssim 10$). After reionisation, it exists mostly in dense self-shielded clumps within galaxies, themselves tracers of the underlying cosmological matter distribution, up to the present day \citep{2001JApA...22..293B, Battye:2004re, 2008PhRvD..78j3511P, 2010ARA&A..48..127M}.

The redshifted 21cm radio line provides the distinctive observational signature of neutral hydrogen gas (\textsc{Hi}) across this entire timespan. It is variously seen in absorption and emission during early epochs, depending on the dominant coupling of the hydrogen gas to other processes, and is mostly seen in emission at late times post-reionisation \citep{2008PhRvD..78j3511P}. The spatial variations of the 21cm signal, as a function of direction and frequency, trace the large-scale structure of the Universe, as well as the ionisation state of the intergalactic medium (IGM) at higher $z$ \citep{1997ApJ...475..429M}.

Mapping out the brightness temperature fluctuations generally requires large radio telescope arrays, with sensitivity on a range of angular and spectral scales \citep{2014ApJ...782...66P, Bull:2014rha, 2024JOSS....9.6501M}. The fluctuating signal is quite different from the much brighter foregrounds, however, which are spectrally smooth except for modulating effects due to the chromatic response of the instruments. The foreground contamination can be isolated into a wedge region in Fourier space \citep{Thyagarajan:2015ewa}, leaving a cleaner `window' in which the 21cm fluctuations should in principle dominate. This permits a `foreground avoidance' approach, in which only the relatively uncontaminated region of Fourier space needs to be retained to make a detection of the 21cm signal.

The sky-averaged, or global, 21cm signal, is a coarser observable, consisting only of a mean brightness temperature as a function of frequency \citep{1999A&A...345..380S}. The spatial fluctuations of the 21cm emission are largely averaged out on solid angles that are a significant fraction of the full sky, leaving only the smooth frequency-dependence of the average signal. This has a series of distinctive peaks and troughs, corresponding to epochs when large-scale heating or cooling of the IGM occurred, or the hydrogen gas coupled to different physical processes that changed its spin temperature \citep{2005MNRAS.363..818S, 2010PhRvD..82b3006P, Glover2014}. The mean brightness temperature is given by
\be
\overline{T}_b(z) \propto \Big (1 - x(z) \Big ) \left ( 1 - \frac{T_{\rm bkgd}(z)}{T_S(z)} \right ) \sqrt{1 + z},
\ee
where $x$ is the mean ionised fraction, $T_{\rm bkgd}$ is the background photon temperature (usually set to the CMB temperature, $T_{\rm bkgd} \approx T_{\rm CMB}$), and $T_S$ is the spin temperature. If the spin temperature is larger than the background temperature, the 21cm line is seen in emission. The proportionality constant depends on cosmological parameters such as the baryon and total matter fractions, and the Hubble parameter.

The evolution of the ionised fraction, and background and spin temperatures, is generally expected to be quite smooth, with peak and trough features in the mean brightness temperature typically spanning widths of $\gtrsim 10$~MHz or so in most physical models \citep{2020MNRAS.495.4845C}. This makes separation of the faint 21cm signal from foreground contamination much harder, as instrumental, terrestrial, and atmospheric artefacts can readily impose structure on similar frequency scales \citep{2021MNRAS.506.2041A, 2021MNRAS.503..344S, 2024MNRAS.528.1945T, 2024ApJ...967...87W, 2024MNRAS.527.2413P}. As a result, extremely accurate instrumental calibration and characterisation is needed to make global signal experiments feasible.

This is the central challenge for global signal experiments, and a variety of approaches have been taken to try and reach the requisite levels of precision. Examples of contemporary global signal experiments include EDGES \citep{2012RaSc...47.0K06R, 2018Natur.555...67B}, SARAS (2 and 3) \citep{2018ExA....45..269S, 2021arXiv210401756N}, REACH \citep{2022NatAs...6..984D}, MIST \citep{2024MNRAS.530.4125M}, and PRIZM \citep{2019JAI.....850004P}, along with several other past, planned, and proposed efforts \citep[e.g.][]{2012RaSc...47.0K06R, 2014arXiv1409.2774P, 2015PASA...32....4S}. These differ in terms of their detailed implementation, but share some common features. First, all of them use compact antennas, either dipoles (e.g. blade dipoles for EDGES and MIST, shaped petal/hexagonal dipoles for PRIZM and REACH respectively), spherical and conical monopoles (SARAS-2 and SARAS-3 respectively), and conical log-spiral (REACH's second antenna). Many of these have been optimised to provide a wideband and relatively frequency-independent response, i.e. the beam full-width at half-maximum (FWHM) is almost constant in frequency \citep[e.g.][]{2022JAI....1150001C}. Most of them also require a ground plane {\edit to suppress spectral artifacts caused by reflections from the ground \citep[although see e.g.][]{2024ApJ...961...56M}. These ground planes sometimes span a wide area or are raised off the ground to avoid edge effects and other complications, but are potentially also a source of residual systematic artefacts themselves \citep{2019ApJ...874..153B, 2022MNRAS.515.1580S}.} In terms of the receiver electronics, these experiments typically use some elaboration of a Dicke-switched design \citep{dicke1946measurement}, in which the system frequently switches between observing the antenna and observing one or more calibration loads. There is more variation in the number of internal calibration targets, and how reflections in the signal chain are measured \citep[e.g. see][]{2024Univ...10..236S}, as well as differences in data transport (e.g. whether RF over fibre links are used for example).

These commonalities are both a manifestation of convergent design (a sign that certain design features work well and so tend to be adopted widely), and a potential risk. Reflections and standing waves have been identified as important possible sources of spurious signals that can mimic the 21cm signal, and it is difficult to fully rid dipoles with ground planes of these issues (although several mitigations have been put in place by the experiments). Sensitivity to the detailed properties of the ground plane, and even what lies beneath it, introduce difficult modelling challenges \citep{2022RaSc...5707558R, 2022MNRAS.515.1580S}. The broad beams of the compact antenna designs result in a general sensitivity to the surrounding environment, including beneath the antenna. While generally seen as a robust calibration strategy, Dicke switching can also have some drawbacks, e.g. in terms of needing rapid switching between different signals paths, which must be separately characterised.

In this paper, we introduce the {\it Remote H$_{\rm I}$ eNvironment Observer} (RHINO), an attempt to build a global signal experiment that has a high level of independence from the `convergent' designs. The central feature of RHINO is that it uses a large horn antenna instead of a compact or resonant design. Horn antennas are well-known for being characterisable with great accuracy \citep{balanis1988horn, bird2007horn}, hence their use in a number of precision-calibrated applications over the years, e.g. for Cosmic Microwave Background experiments, plus lower frequency experiments like TRIS \citep{2008ApJ...688...12Z} and L-BASS \citep{Zerafa}. With some tuning of the design, they can also achieve low sidelobes, a wideband response, and low cross-polarisation. Since the horns themselves are essentially waveguides, rather than `active' receiving surfaces, there is also an advantage in terms of shielding of the actual feed from the surrounding environment, including the ground, and there is no strict need to incorporate a ground plane.

In another choice to promote design independence, we also include a continuous wave (CW) calibration source \citep{2019MNRAS.489..548P}. This injects a known signal of large amplitude but very narrow spectral width, and permits continuous monitoring during observations without needing to Dicke switch. This can be used to significantly suppress correlated gain fluctuations if they are sufficiently correlated in frequency.


The purpose of this paper is to describe the RHINO horn antenna (Sect.~\ref{sec:horns}) and system designs (Sect.~\ref{sec:system}), 
and provide an overview of the continuous wave calibration approach (Sect.~\ref{sec:cal}). We also perform forecasts for how well physical 21cm global signal parameters would be recovered in idealised scenarios (Sect.~\ref{sec:forecasts}), and conclude with a brief report on the current status of a prototype system currently under construction at Jodrell Bank Observatory (Sect.~\ref{sec:prototype}). The basic properties of the design concept are summarised in Table~\ref{table:properties}.

\begin{table}
\centering
  \begin{tabular}{ll}
  \hline
  {\bf RHINO system concept} \\
  \hline
  Freq. range (minimum)   & 65 -- 80 MHz \\
  Freq. range (target) & 60 -- 85 MHz \\
  Beam FWHM (65 -- 80 MHz)  & 44$^\circ$ -- 35$^\circ$\\
  \hline
  Horn shape & Pyramidal \\
  Horn aperture & $7.20 \times 6.00$m \\
  Flare section height & 4.30m \\
  Waveguide aperture & $3.54 \times 2.00$m \\
  Waveguide height & 2.98m \\
  \hline
  Site              & Jodrell Bank Observatory (UK) \\
  Latitude     & $53^\circ14^\prime10^{\prime\prime}$N \\
  Longitude    & $2^\circ 18^\prime 26^{\prime\prime}$W \\
  Altitude (above sea level)     & 77m \\
  \hline
  \end{tabular}
  \caption{Basic properties of the RHINO system concept in its current iteration. Alternative horn geometries and sites are being explored however.} \label{table:properties}
\end{table}

\section{Horn antenna design} \label{sec:horns}

In this section, we briefly review the basic principles of horn antenna design. We then set out the design requirements for the RHINO antenna, and describe a suitable candidate design that can achieve the large size of horn required for operation at around 70~MHz while maintaining cost effectiveness. Finally, we present a set of electromagnetic simulations of the candidate design and study the frequency and spatial dependence of the simulated antenna patterns.

\begin{figure*}
    \centering
    \includegraphics[width=\columnwidth]{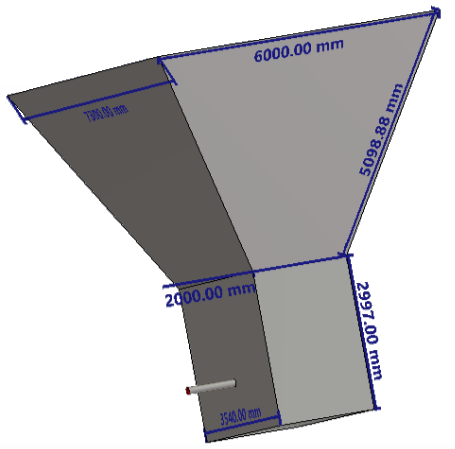}
    \includegraphics[width=\columnwidth]{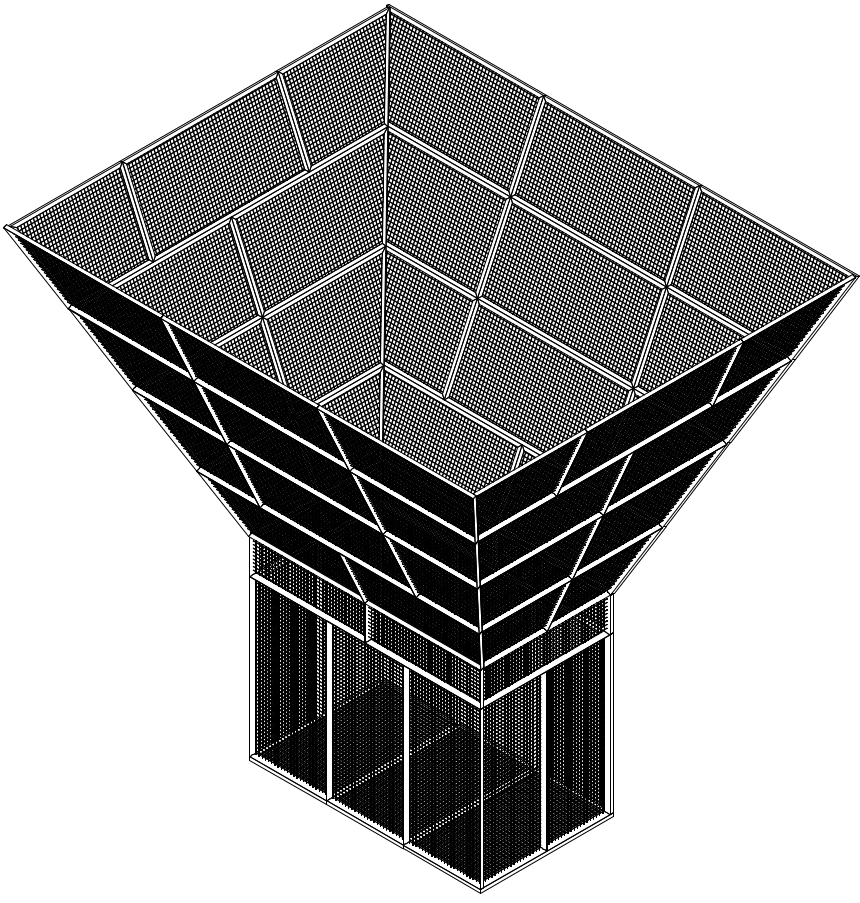}
    \caption{\edit {\it (Left):} Dimensions of the pyramidal horn that was selected as the reference design. The flare section has an aperture of $7.3 \times 6.0$m, an edge slant length of 5.099m, and an overall vertical height of 4.3m. The rectangular waveguide section has an aperture of $3.54 \times 2.00$m and a height of 2.997m. The horn will be static, pointing at zenith, and will have a welded mesh conductive surface. The location of the feed is marked on the waveguide. {\it (Right):} Detailed design drawing showing the multi-panel construction, consisting of $2.4 \times 1.2$~m mesh panels cut to size/shape and welded into steel angle frames. These are then bolted together to form a self-supporting structure.}
    \label{fig:horn}
\end{figure*}

\subsection{Basic principles} \label{sec:horns:basic}

Horn antennas can be thought of as tapered waveguides that shape the incoming EM radiation in such a way that the impedance of the receiver is smoothly matched to the impedance of free space. The horn itself merely provides a spatially-varying conductive boundary, and is not itself an active receiving surface; a simple coaxial feed set far back into the waveguide is what actually absorbs the EM waves. As with waveguides, only certain discrete modes satisfy the boundary conditions within the horn, and one can analyse different horn geometries according to which modes are supported \citep{balanis1988horn}.

The minimum size of a horn is first set by the cutoff frequency, $\nu_c$, of its terminal waveguide. For a rectangular waveguide to support a mode ${\rm TE}_{mn}$, the cutoff frequency is defined as \citep{bird2007horn}
\be
\nu_c = \frac{c}{2} \sqrt{\left (\frac{m}{D_H}\right)^2 + \left (\frac{n}{D_E}\right)^2},
\ee
where $m$ and $n$ label the order of the modes in the H and E planes, and $D_H$ and $D_E$ are the linear dimensions of the waveguide cross-section in the H and E plane directions respectively. As an example, to support the lowest order ${\rm TE}_{10}$ mode at 50~MHz and above, a waveguide side length of at least 3m is required, and so any flared horn section attached to this will necessarily have a wider aperture than this.

The next consideration is the desired full-width at half-maximum (FWHM) of the antenna pattern at the low end of the frequency band, set roughly by the diffraction limit, $\Delta \theta \sim\lambda / D$, for an aperture of diameter $D$. At 60~MHz ($\lambda \approx 5$m), this implies a diameter of around 7m to achieve a $40^\circ$ beam FWHM. 

For a pyramidal (rectangular) horn of a given aperture diameter, it is possible to find a so-called `optimum horn' with dimensions that maximise the antenna gain. The optimum H and E-plane aperture side lengths are related to the height of the horn flare section, $h$, by \citep{balanis1988horn}
\be
D_H = \sqrt{3 \lambda h};~~~~~~D_E = \sqrt{2 \lambda h}.
\ee
The optimum is only achieved at a particular wavelength. Setting the target frequency to 70~MHz ($\lambda \approx 4.3$m), for an H-plane side of length 7.2m, this implies $h=4$m (or 6m for the E-plane). The horn length also affects other properties of the antenna pattern. The curvature of the wavefront is changing as it propagates between aperture and waveguide, and if the flare section is not long enough, some curvature will remain across the aperture, leading to an appreciable phase error and (e.g.) increased sidelobes. Diffraction at the edge of the aperture also contributes to an increased backlobe and far sidelobes. At high frequencies, design features such as corrugations can be used to more gently taper the field and improve the antenna pattern shape, but these would be difficult and costly to manufacture into a very large horn. Other possibilities, such as a choke around the aperture to suppress diffraction, are likely to be more practical in the $\lesssim 100$~MHz range.

\subsection{Design requirements} \label{sec:horns:designs}

The following design requirements were identified for the antenna:
\begin{enumerate}[label=(\roman*),leftmargin=0.5cm]
  \item Beam FWHM between $25^\circ - 45^\circ$ between 65 -- 80 MHz.
  \item Return loss better than $10$~dB between 65--80~MHz.
  \item First sidelobes below $-$20~dB (peak-normalised power).
  \item Backlobe below $-$30~dB (peak-normalised power).
\end{enumerate}
{\edit These requirements were set on a qualitative basis, rather than flowing in detail from the scientific objective of suppressing systematic contamination of the recovered 21cm signal. The latter would involve end-to-end simulations that are strongly contingent on a choice of foreground removal method for instance \citep[although such design studies have been attempted for other experiments, e.g.][]{2022MNRAS.509.4679A}. Nevertheless, it is useful to have a target for optimisation. We justify each requirement as follows.} The beam FWHM range was chosen to ensure a broad instantaneous field of view, so as to fairly sample the monopole on the sky while limiting the sensitivity to low elevations, where most difficult-to-model terrestrial emission is found. The return loss requirement defines the minimum desired bandwidth, from 65--80~MHz (21\% fractional bandwidth). We target a broader band of 60--85~MHz (34\% fractional bandwidth), but anticipate degraded sensitivity at the band edges due to the receiver bandpass (which includes an FM bandstop filter for example). This band corresponds to redshifts between 15.7 -- 22.7, and was chosen to overlap substantially with the EDGES feature, without extending into the FM band. Extending the band to lower frequencies would result in a larger and increasingly difficult to manufacture structure. We study the impact of targeting a relatively small bandwidth on recovery of 21cm global signal model parameters in Sect.~\ref{sec:forecasts}.

\begin{figure}
    \centering
    \includegraphics[width=1\columnwidth]{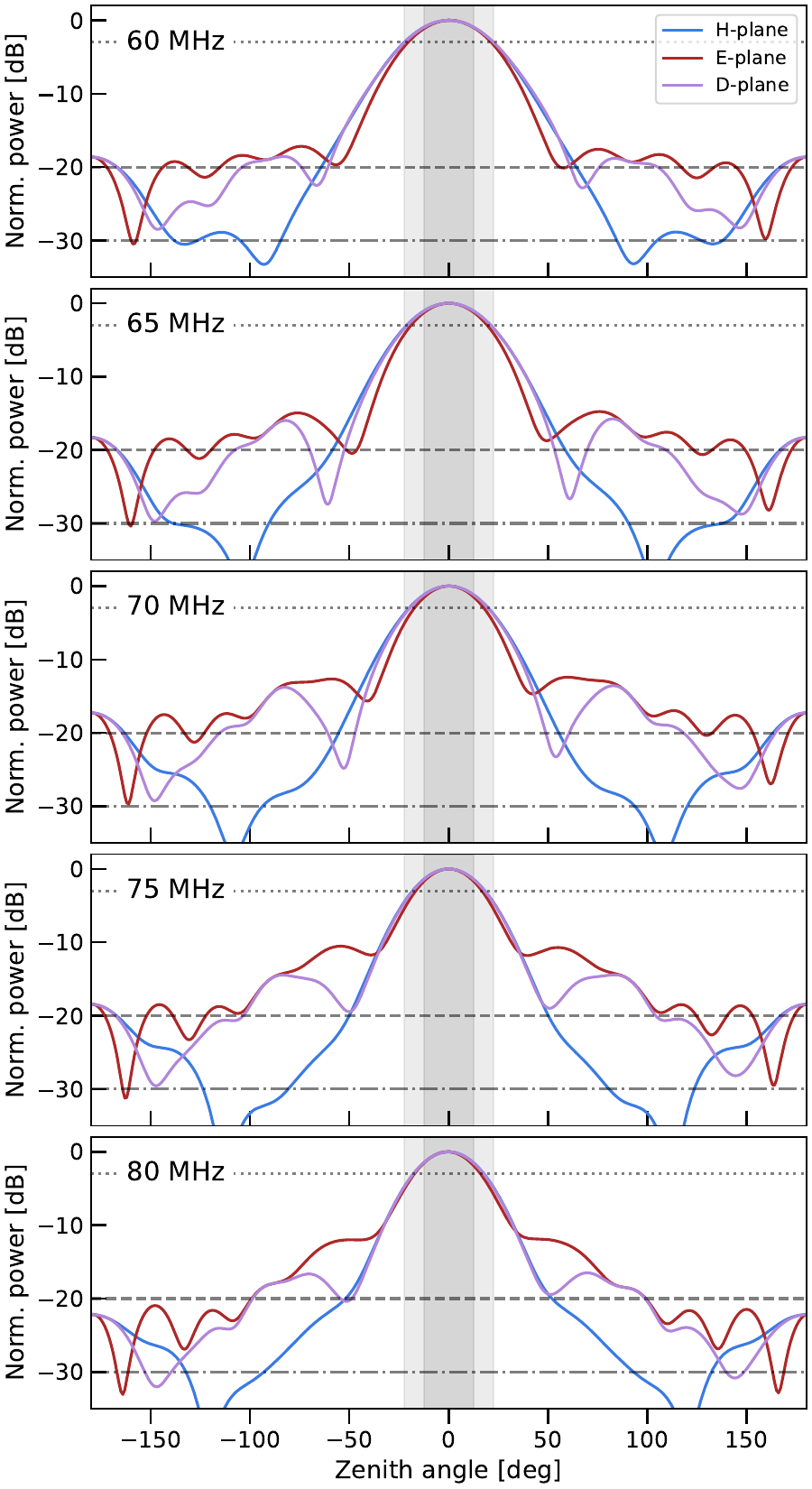}
    \caption{Slices through the normalised antenna pattern, as a function of frequency. The horizontal lines show: (dotted) the $-3$~dB (half-maximum) point; (dashed) the target sidelobe level; and (dot-dash) the target backlobe level. The shaded vertical regions show the target beam FWHM at the bottom and top of the design band.}
    \label{fig:beams1d}
\end{figure}

Sidelobes are typically more chromatic than the mainlobe, and are harder to measure in the field. As such, sidelobes are a potentially important and uncertain source of spectral variation in the data. The sidelobe level target was chosen to try to keep them below 1\%, so that beam modelling errors would be correspondingly less important (although not necessarily negligible). At higher frequencies, it has been possible to build horn antennas with very low sidelobe levels, approaching and even surpassing $-30$~dB \citep[e.g.][]{1966ITAP...14..605L, 1993ITAP...41..357Y, 2008ApJ...688...12Z, Wuensche:2021dcx}. This has been achieved through designs with longer flare sections (which helps reduce phase errors) and the use of internal corrugations to shape the field within the horn. While technically feasible at lower frequencies too, either of these options would add significant expense and practical difficulties in the manufacture and maintenance of the antenna.\footnote{For concreteness, our goal is to restrict the materials and construction costs to {\edit substantially below US\$100k at 2025 prices.}} On the other hand, by limiting ourselves to simpler and relatively compact geometries, we can expect generally larger sidelobe levels. We note that the first sidelobe is the main focus here; as well as normally being the largest, for the large beamwidths we consider, it is also likely to be the only sidelobe above the horizon.

Finally, the backlobe requirement is intended to minimise the sensitivity of the antenna to the properties of the ground. As discussed above, dependencies on the ground plane shape and size, and even the electrical properties of the underlying soil, can prove challenging to model for compact antennas such as dipoles \citep{2022MNRAS.515.1580S, 2025MNRAS.538.1301P}. For horn antennas, the backlobe is mostly due to diffraction \citep[e.g. see][]{1966ITAP...14..605L, clarricoats1984corrugated}, and so can in principle be managed through modifications of the antenna optics, e.g. by adding chokes, corrugations, or screens at the aperture. More complex solutions such as loading the horn with a metamaterial have also been proposed \citep[e.g.][]{qi2015suppressing, polo2018compact}.

\begin{figure}
    \centering
    \includegraphics[width=1\columnwidth]{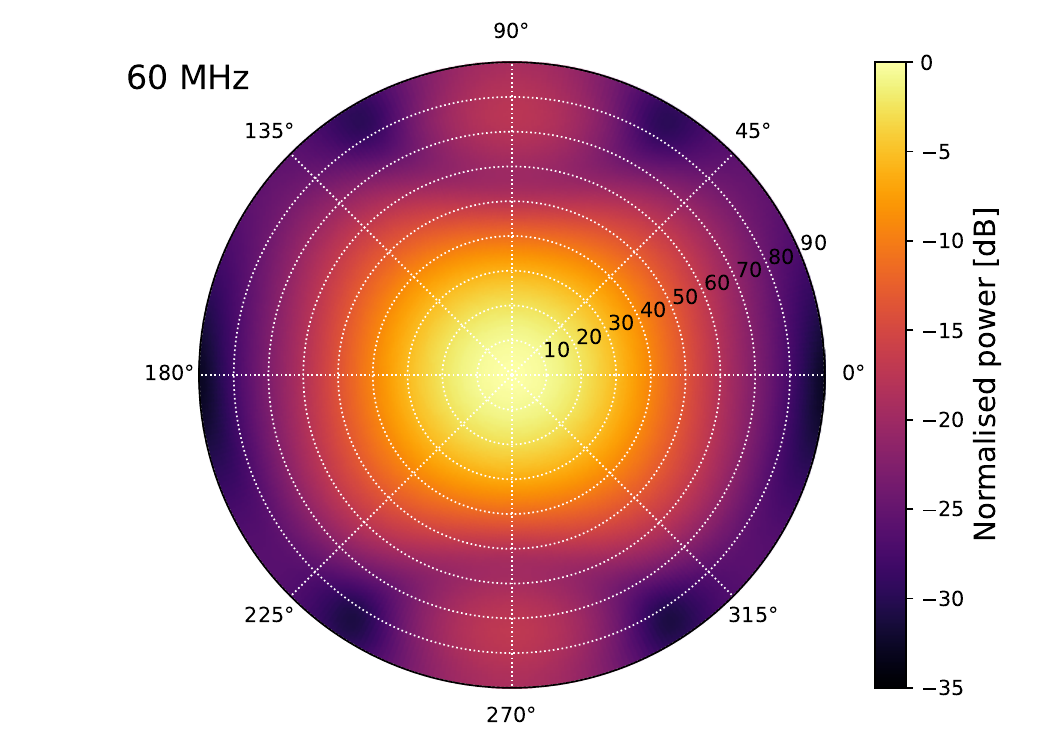}
    \includegraphics[width=1\columnwidth]{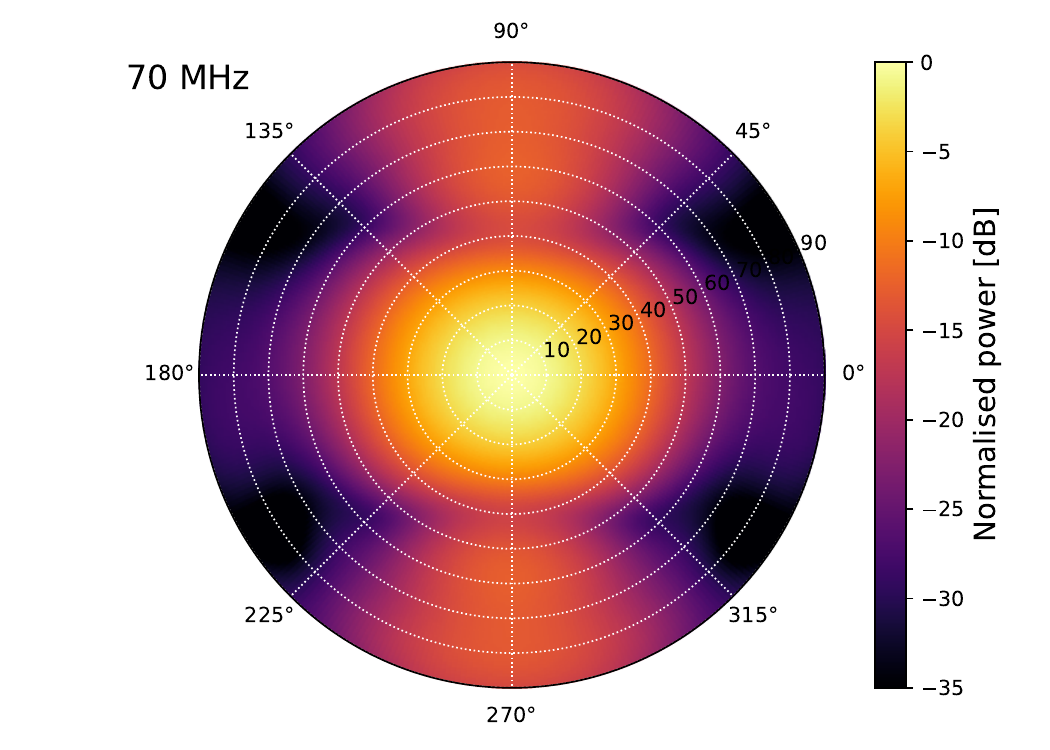}
    \includegraphics[width=1\columnwidth]{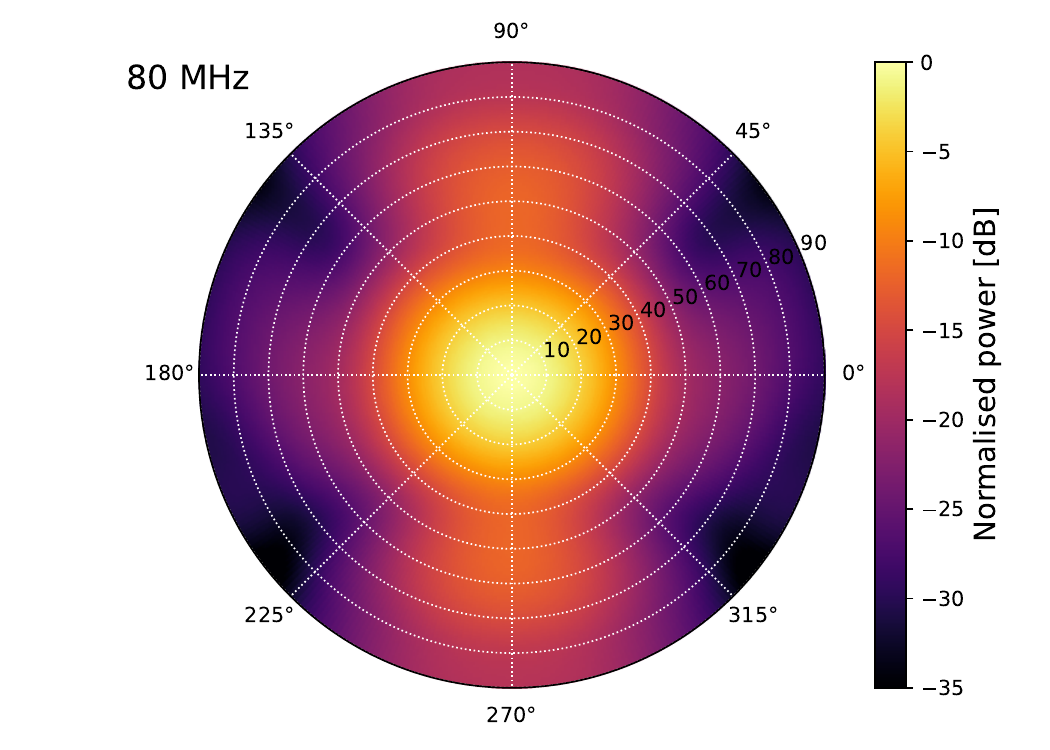}
    \caption{Normalised antenna patterns as a function of frequency, for angles above the horizon.}
    \label{fig:beam}
\end{figure}

\begin{figure}
    \centering
    \includegraphics[width=\columnwidth]{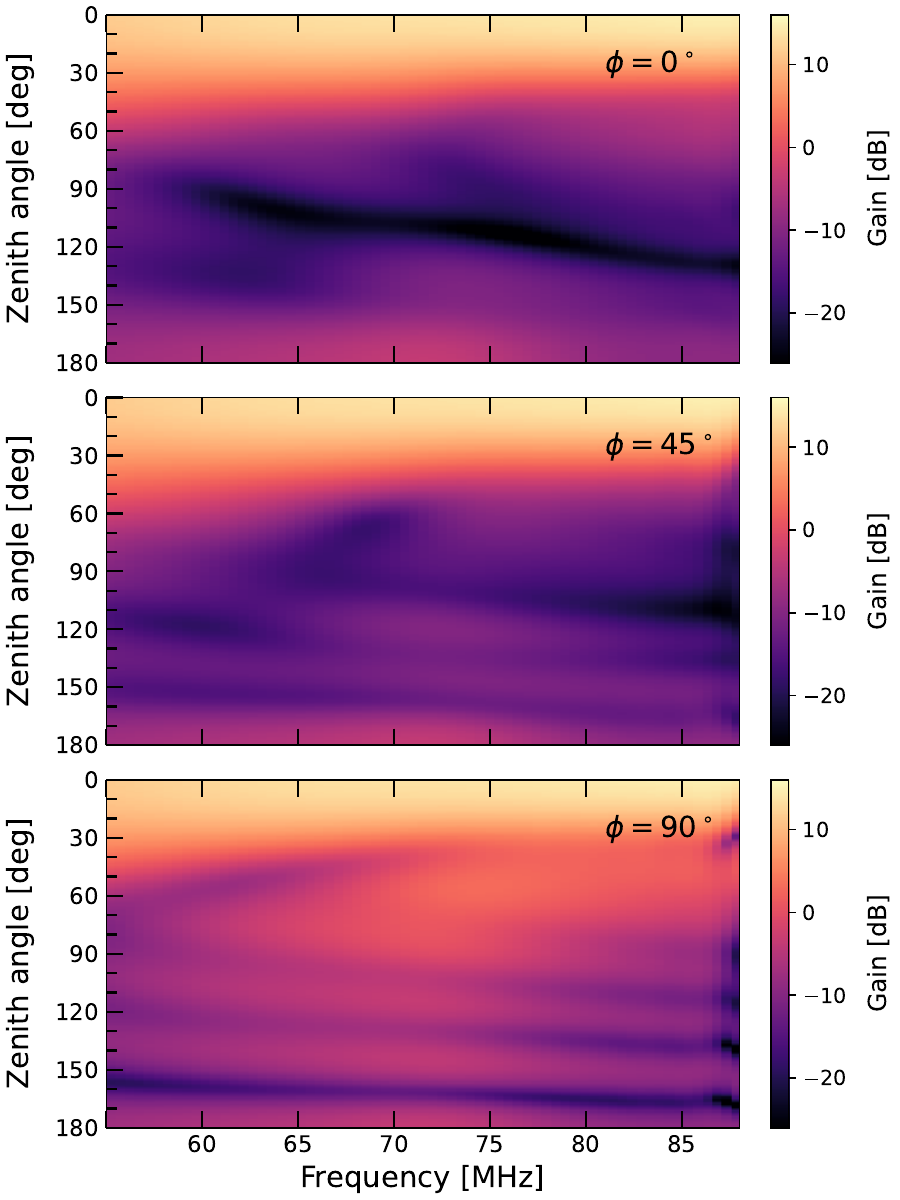}
    \caption{Unnormalised antenna pattern as a function of frequency and zenith angle, for different slices in azimuth ($\phi)$.}
    \label{fig:beamfreqs}
\end{figure}

\subsection{Electromagnetic simulations} \label{sec:horns:em}

{\tt CST Studio}\footnote{\url{https://www.3ds.com/products/simulia/cst-studio-suite}} was used to run full-wave electromagnetic simulations of the horn antenna. In the first instance, the horn was modelled with perfect electrical conductor, standing in free space (no ground plane), with a $50\Omega$ coaxial input port feeding the antenna. Several optimisation loops were run to try to find a simple pyramidal horn design that satisfied the requirements above while also reducing the linear dimensions of the horn as much as possible. The waveguide length and feed placement were particularly consequential in achieving the required bandwidth.

\begin{figure}
    \centering
    \includegraphics[width=\columnwidth]{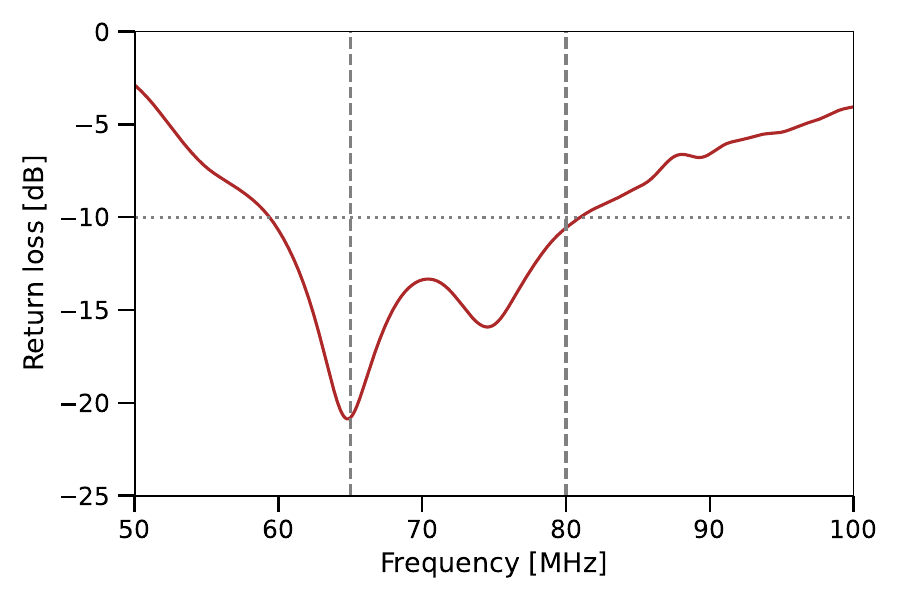}
    \caption{Return loss ($S_{11}$) of the antenna as a function of frequency. The design band is marked by vertical dashed lines, and the target return loss within this band is marked by a dotted line.}
    \label{fig:s11}
\end{figure}

Following this process, we arrived at the reference design shown in Fig.~\ref{fig:horn}. Antenna patterns are shown in Figs.~\ref{fig:beams1d}, \ref{fig:beam}, and \ref{fig:beamfreqs}. The horn aperture is $7.3 \times 6.0$m, with a flare length (along the corner of the flare section) of 5.1m, for a flare section vertical height of 4.3m. The waveguide has an aperture of $2.0\times 3.5$m and a vertical height of 3m. This has a gain of 12.5~dB and FWHM of $44^\circ$ at 65~MHz (Figs.~\ref{fig:beams1d} and \ref{fig:beamfreqs}), and achieves a return loss of $S_{11} < -10$~dB across the design band (Fig.~\ref{fig:s11}). The (relative) backlobe level at 65~MHz is $-18.4$~dB however, and the sidelobe level is around $-15$~dB, both of which are outside the design requirements. This behaviour is expected; in a rectangular horn antenna, the electric field distribution over the E-plane walls is uniform, and as a result, large sidelobe levels in the far-field radiation patterns of the corresponding planes will be observed \citep{kraus2002antennas} unless the E-field distribution can be modified.

The antenna patterns are plotted at 10~MHz intervals in Fig.~\ref{fig:beams1d} and Fig.~\ref{fig:beam}, and as a function of frequency and zenith angle in Fig.~\ref{fig:beamfreqs}. While the mainlobe FWHM varies smoothly with frequency, there is significant spectral structure in the sidelobes. This could be controlled somewhat by using a more complex horn geometry, but this is not the main focus of the RHINO design. Instead, we prefer a simpler design that can be simulated and characterised very accurately, even if the antenna pattern itself is ultimately more complicated. We seek to demonstrate the benefits and disadvantages of this approach in future work, including the construction of a scaled-down prototype horn antenna, and more sophisticated EM simulations.

To recap, the aim of this design exercise was to find the simplest horn antenna that could approach the requirements above. While a simple and workable design was found, it falls outside the target requirements for the sidelobe and backlobe levels, and so in practice a refined design is needed. Several elaborations are possible, and will be explored through detailed simulations in a future work. Briefly, these include the addition of choke rings at the aperture (to suppress the backlobe), and a variable flare angle and/or corrugations around the transition from the flare section to the aperture, to improve matching, and hence reduce sidelobes and improve bandwidth. A densely corrugated design is likely to add significant additional complexity and expense for {\edit a horn approaching 10m in size (see Sect.~\ref{sect:corrugatedhorn}), and so we have not adopted this option, although we note that corrugated designs have been demonstrated for horns of a couple of metres in size used by the BINGO experiment \citep{Wuensche:2021dcx}.} Other options include the construction of an extensive ground screen, to suppress emission from low elevations and below the horizon, as well as more sophisticated optics such as lenses.


\subsection{Sensitivity to ground conditions} \label{sec:horns:ground}

One of the main motivations for using a horn antenna instead of a compact antenna (such as a dipole) is that horns are expected to be less sensitive to the detailed conditions of the ground or ground plane, making it less important to model these components with high accuracy. In this section, we present illustrative EM simulations to examine this expectation, with a more detailed investigation to be presented in Elmakadema et al.
 ({\it in prep.}).
 
\begin{figure}
    \centering
    \includegraphics[width=\columnwidth]{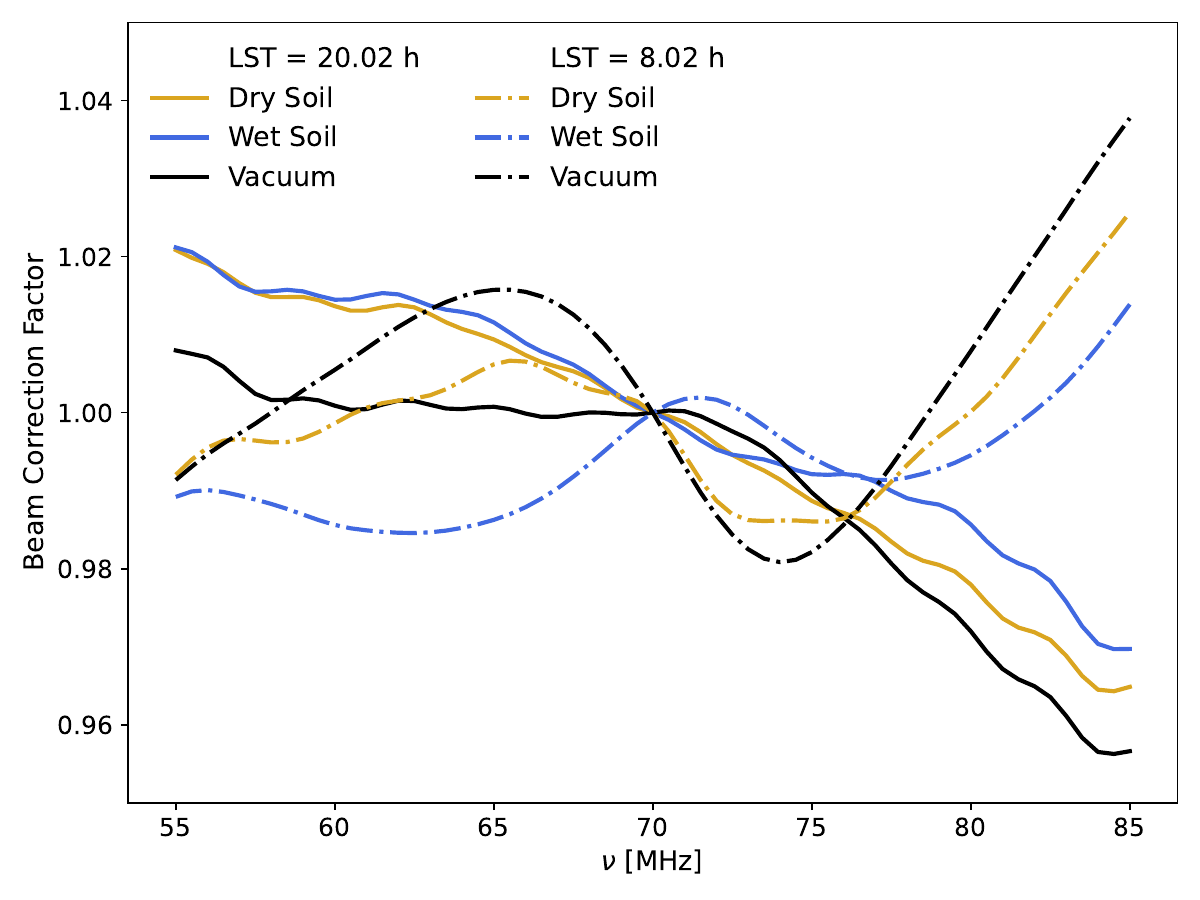}
    \caption{Beam correction factor as a function of frequency, for a single reference time, for three different assumptions about the ground conditions beneath the horn antenna: vacuum (no ground); uniform dry soil ($\rho = 0.001$\,S\,m$^{-1}$, $\epsilon_r = 2$); and uniform wet soil ($\rho = 0.1$\,S\,m$^{-1}$, $\epsilon_r = 7$).}
    \label{fig:beamcorr}
\end{figure}

To study the sensitivity to ground conditions, we use the same horn antenna design as in Sect.~\ref{sec:horns:designs}, but alter the boundary conditions to represent three different idealised ground scenarios -- vacuum (the same as the models shown in the previous section), wet soil, and dry soil. For the non-vacuum cases, the EM simulation domain is a hemisphere of radius 20m immediately beneath the bottom edge of the waveguide, with a uniform conductivity and dielectric constant representative of either wet or dry soil, as reported in \citet{2022MNRAS.515.1580S}. These conditions are intended to bracket reasonable wet and dry conditions, with `typical' soil conditions falling in between depending on the site. Convergence tests were performed to ensure that the choice of domain and mesh parameters in the simulation are appropriate \citep[c.f.][]{Mahesh:2021rly}.

Next, we calculate the beam correction factor for each scenario, using the reprocessed Haslam 408\,MHz map \citep{Haslam1982, 2015MNRAS.451.4311R} scaled by a synchrotron power-law spectrum as our sky brightness temperature model. The beam correction factor is calculated as defined in \citet{2022MNRAS.515.1580S},
\be
B_{\rm corr}(\nu, t) = \frac{\int _{\Omega}T_{\rm sky}(\nu_{\rm ref}, \hat{n})B(\nu, t, \hat{n})d\Omega }{\int_\Omega T_{\rm sky}(\nu_{\rm ref}, \hat{n})B(\nu_{\rm ref}, t, \hat{n})d\Omega} 
  \frac{\int_{\Omega}B(\nu_{\rm ref}, \hat{n})d\Omega}{\int_{\Omega} B(\nu, \hat{n})d\Omega}.
\ee
Here, $B(\nu, t)$ is the power beam pattern as a function of frequency and local sidereal time of observation, $t$, and $T_{\rm sky}$ is the sky brightness model. A reference frequency of $\nu_{\rm ref} = 70$~MHz has been chosen in what follows. This factor accounts for the frequency dependence of the mapping between antenna temperature and brightness temperature, and the application of this correction is the main part of a 21cm global signal analysis where a precise beam model is needed. The correction factor is relatively spectrally smooth for dipole-type antennas \citep[e.g. see][]{2021MNRAS.506.2041A}, which have antenna patterns dominated by a mainlobe that is relatively stable with frequency. Interactions between the antenna and its ground plane or the ground itself (for example) can contribute additional chromaticity however \citep{2019ApJ...874..153B, 2022MNRAS.515.1580S}. Horns and other aperture antennas, on the other hand, are intrinsically more chromatic, and tend to have more complex sidelobes, but do not conventionally require a ground plane to be incorporated in the design. Most of their sensitivity to the ground comes through far sidelobes caused by diffraction around the aperture, which can be mitigated through design choices to minimise sidelobes for example.

Fig.~\ref{fig:beamcorr} shows the beam correction factor for the three scenarios at two local sidereal times: LST = 20h, when a cold region of the sky is at zenith, and LST = 08h, when the Galactic plane is transiting. Focusing first on the 20h case, we see that the beam correction factor changes relatively little between the wet and dry cases, which supports the expectation that the horn antenna design should be insensitive to ground conditions. There is a larger difference between these cases and the vacuum case, as the change in boundary conditions now allows for reflections from the ground that are not possible in vacuum. For the 08h case, when the bright Galactic plane is overhead, the differences between the three cases become more pronounced, but are still at the 1\% level or less. This is comparable to the results of \citet{2022MNRAS.515.1580S} (see their Fig.~11), which shows beam correction factor variations of around $\pm 0.3\%$ between the wet and dry cases for different shapes and sizes of ground plane (in this case, for a LEDA triangular dipole antenna).

\subsection{Mechanical design} \label{sec:horns:fab}

In addition to establishing the expected performance of a given antenna design through electromagnetic modelling, it is also important to consider its mechanical feasibility. A structure of dimensions $7.3 \times 6.0 \times 8.1$~m is not particularly large by modern engineering standards, but it is nevertheless helpful to identify ways of achieving a given electromagnetic performance while reducing manufacturing costs and complexity.

Given the long wavelengths involved, the surface accuracy requirements are likely to be relatively mild (although this accuracy must be achieved over a large surface area). A reasonably stringent target surface accuracy of $\lambda / 80$ equates to about 5~cm at 70~MHz. Depending on the type of surface, distortions due to the changing ambient temperature and wind loading may approach this level, perhaps necessitating additional structural reinforcement to stiffen the surface.


The conductivity of the surface is also important, as Ohmic losses will reduce the antenna efficiency. \cite{Findlay1990} used welded sheet metal to minimise the losses in a pyramidal horn antenna with aperture $5.4 \times 4.0$m and slant length 36.7m for example. For a large vertical antenna, however, this presents challenges in terms of the necessary support structure, as well as wind loading, thermal expansion etc. An alternative is to use welded mesh, which has benefits in terms of reduced cost and weight, but will have higher losses. It is also less straightforward to achieve suitable conductive joins between pieces of mesh, as the narrow wire is relatively fragile for welding. {\edit To overcome this, we selected 8g (4mm diameter) mild steel welded mesh panels with $50 \times 50$~mm mesh aperture. Each $2.4 \times 1.2$~m mesh panel is cut to size and then welded into a box frame made from $50 \times 50 \times 5$~mm steel angle. The mesh gauge is sufficiently thick to permit efficient direct welds onto the angle. The panels can then be bolted together so that the flat angle surfaces perpendicular to the mesh are directly in contact. Fig.~\ref{fig:horn} shows a detailed design for this arrangement, consisting of 59~panels for a total mass of around 1800~kg. This structure is self-supporting, but requires bracing with guy lines or similar to withstand wind loading etc.}


As a final note, alternative construction methods were also considered, including a horn-reflector design (a horizontal horn coupled to a section of parabola), and a vertical horn consisting of wire loops strung between telegraph poles. The former would permit a longer horn to be built without requiring as much vertical construction, but has higher material costs, while the latter minimises costs but offers less control over surface accuracy.

\begin{figure}
    \centering
    \includegraphics[width=1\columnwidth]{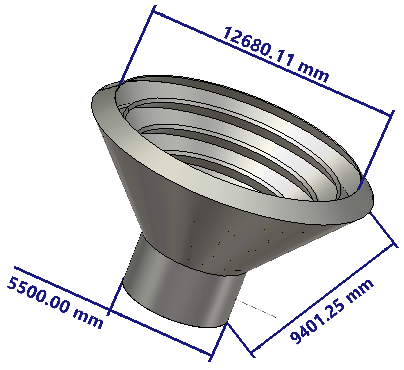}
    \caption{Dimensions of the proposed corrugated conical horn design. There are five corrugations in the flare section. Note the significantly larger dimensions than the pyramidal horn; further optimisation is underway to achieve comparable results with reduced linear dimensions.}
    \label{fig:corrugated_diagram}
\end{figure}

\subsection{Alternative corrugated conical horn design}\label{sect:corrugatedhorn}

Another option is to design a corrugated horn, which provides many advantages compared to a conventional pyramidal horn, at the cost of a more complicated design. {\edit We present a suitably optimised alternative design here to illustrate the differences in performance and complexity of construction, and to provide a reference future development path for an enhanced low-frequency absolutely-calibrated horn antenna experiment. Fig.~\ref{fig:corrugated_diagram} shows an alternative horn geometry,} consisting of a conical horn with a small number of large corrugations (five in this iteration). This should be contrasted with more conventional corrugated designs at higher frequencies, which place many corrugations over along the flare section of the horn. Some rules of thumb for corrugation depth and spacing are discussed in \cite{1966ITAP...14..605L} for instance.

\begin{figure}
    \centering
    \includegraphics[width=1\columnwidth]{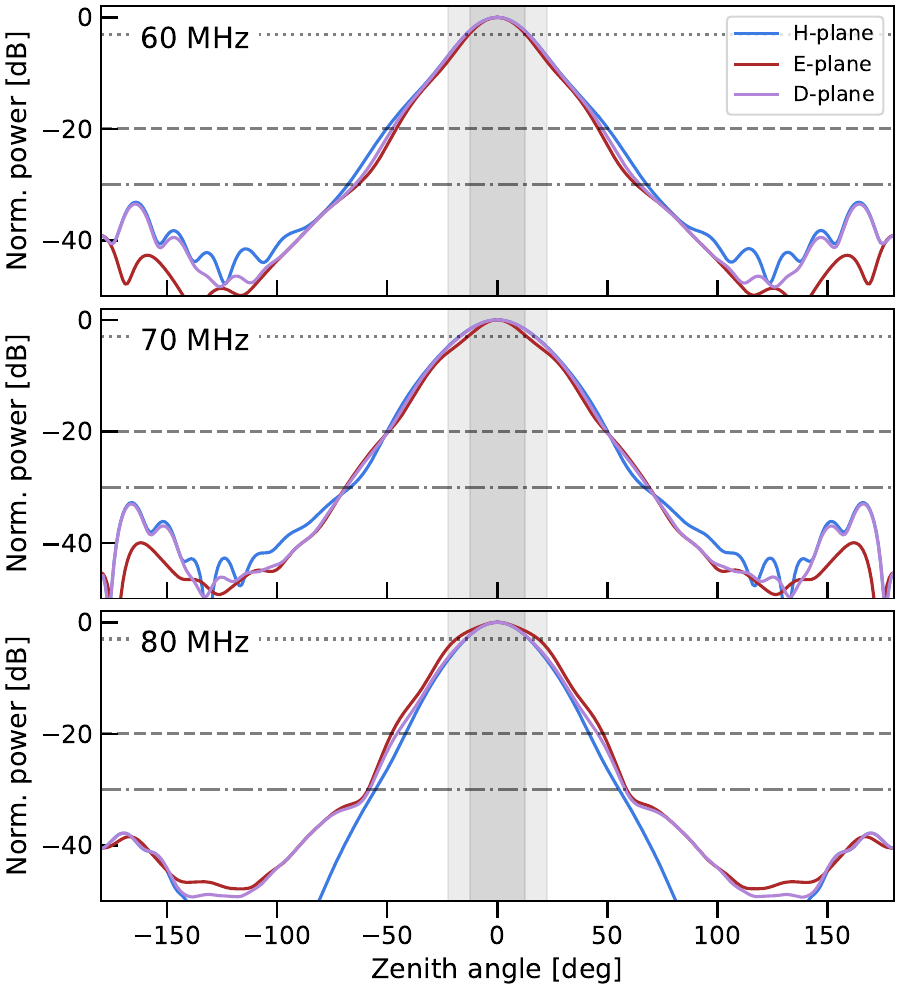}
    \caption{Slices through the normalised antenna pattern as in Fig.~\ref{fig:beams1d}, but now for the corrugated conical horn {\edit shown in Fig.~\ref{fig:corrugated_diagram}}. The sidelobe level and degree of symmetry are significantly improved in this design.}
    \label{fig:corrugated_beams}
\end{figure}

These corrugations are nevertheless successful in significantly improving the antenna pattern, to the extent that it now satisfies all of the design requirements in Sect.~\ref{sec:horns:designs}. Slices through the antenna patterns are shown in Fig.~\ref{fig:corrugated_beams}. The most noticeable improvements are in terms of the sidelobe level, which is now better than $-30$~dB in both the E-plane and H-plane, with a high degree of symmetry. The backlobe is also greatly reduced, and is better than $-30$~dB across the frequency range.

In addition, corrugated designs offer potentially much broader bandwidths, and better cross-polarisation discrimination. However, corrugated horns are much more complex structures compared to a conventional pyramidal horn, and are hence more challenging in terms of their design and construction. Note that the dimensions of the horn shown in Fig.~\ref{fig:corrugated_diagram} are also substantially larger than for the pyramidal horn from Fig.~\ref{fig:horn}; the aperture diameter is 12.7m for instance, versus 7.3m for the pyramid. Further improvements to the design are possible, with the aim of reducing its complexity by reducing the overall size and (possibly) the number of corrugated sections. Given the long wavelengths involved, it may also be possible to approximate the circular profile of the horn with simpler polygonal shapes, e.g. an octagonal profile, without significantly degrading the antenna pattern. This would help improve the feasibility of the construction without driving up the cost, while still achieving the performance benefits of the corrugated and (approximately) conical design.

\section{System design} \label{sec:system}

In this section, we introduce the RHINO system design, which currently exists in a prototype form. The basic system properties are summarised in Table~\ref{table:properties}, and a block diagram for the prototype is shown in Fig.~\ref{fig:sysdiagram}.



\subsection{Enclosure and feed from antenna}
The receiver electronics are placed in a shielded box immediately next to the waveguide section of the horn, with a short shielded co-axial cable connecting the feed to the receiver. The proximity is intended to reduce the impact of cable reflections and pickup in the cable by requiring only very short cable lengths. For coaxial cable with a propagation velocity of $c_{\rm coax} \approx (2/3) c$, avoiding reflections with a characteristic scale of $\Delta \nu \lesssim 100$~MHz requires cable lengths of $L \lesssim c_{\rm coax} / (2 \Delta \nu) \sim 1$~m or less.

\subsection{Warm/hot calibration loads} \label{sec:WarmHotLoads}

In order to provide an absolute reference for the calibration, we incorporate two thermally-stabilised loads into the system. One is kept at $T \approx 110^\circ{\rm C}$ (380~K), while another is heated to $T \approx 220^\circ{\rm C}$ (490~K). These temperatures, and the difference between them, is relatively low compared to the sky, which can be of order a few thousand Kelvin at 70~MHz. The loads are standard $50\Omega$ terminations embedded in small aluminium blocks with dimensions $49 \times 42 \times 42$~mm. We attach thermostatic positive temperature coefficient (PTC) heating pads to one face of each block. After a warm-up phase, the PTC pads self-regulate their temperature when kept at a constant voltage.

To monitor the temperature in real time, Pt-1000 platinum resistance thermistors are also embedded in the aluminium blocks, and are connected to the control system via Analog Devices MAX31865 amplifier boards.\footnote{\url{https://www.analog.com/media/en/technical-documentation/data-sheets/max31865.pdf}} This type of thermistor has response times of $\sim 0.1$s, low self-heating, and typical rms accuracy of $\sim 0.1^\circ{\rm C}$. The temperature probes can be polled rapidly in order to measure temperature changes on sub-second timescales. The loads are housed within an insulated, RF-shielded enclosure that is attached to the outside of the receiver box.


\begin{figure}
    \centering
    \includegraphics[width=1\columnwidth]{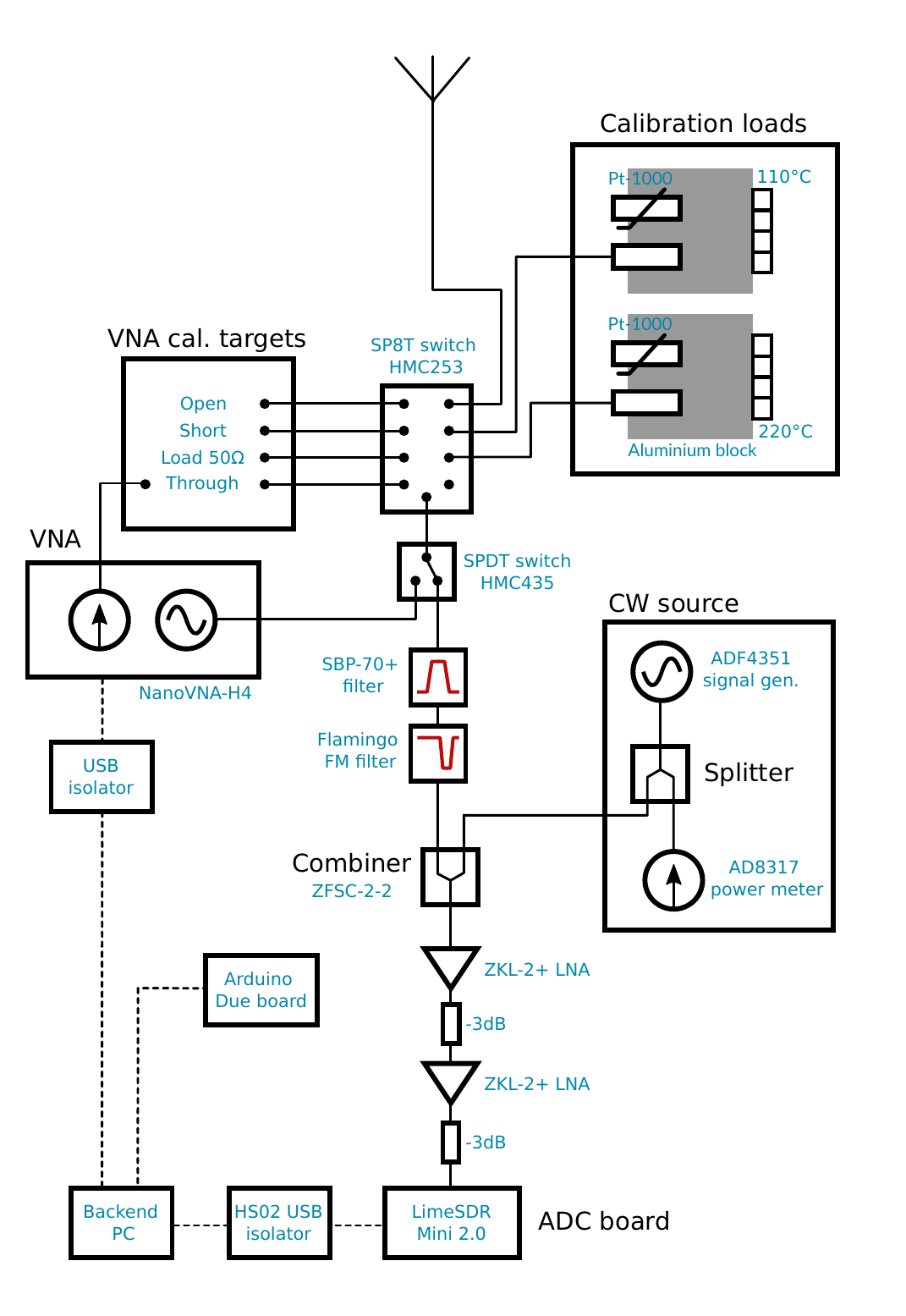}
    \caption{System diagram showing the major components of the receiver, including model numbers for the components used in the prototype system. Solid lines denote coaxial connections via SMA connectors. Dashed lines denote USB connections. Power lines, and control lines from the Arduino board, are not shown.}
    \label{fig:sysdiagram}
\end{figure}

\subsection{Continuous wave source}

One of the novel aspects of our calibration scheme (as far as ground-based 21cm global signal experiments are concerned)\footnote{Continuous wave or `tone' calibration methods have previously been used in a number of related contexts \citep[e.g.][]{bale2016fields, 2021arXiv210305085B}.} is the inclusion of a continuous wave (CW) source. This injects a constant tone into the system that has a large amplitude but only occupies a single frequency channel, leaving most of the band unaffected, and barely increasing the total power into the amplifier. The CW source is monitored in real-time by a logarithmic power meter to track its stability, and can be tuned to different frequencies on demand. For the prototype system, we use an Analog Devices ADF4351 (square wave) synthesiser board attached to an AD8317 power meter via a splitter. The CW module is then connected to the receiver system via a combiner, which sums the CW and antenna signals.

The CW source is intended to provide a continuous time-dependent gain calibration without needing to switch in the calibration loads frequently. Switching changes the power levels into the amplifier and introduces different signal paths with different reflection coefficients, potentially reducing the stability of the system. The CW source, on the other hand, is always present during normal observing of the sky signal, and so drifts in the overall gain level due to $1/f$ noise can be tracked at all times, without any switching required. The addition of another signal path via the combiner does make reflection measurements more complicated however.

This also relies on the bandpass being stable on relatively long timescales, i.e. the correlated gain fluctuations are assumed to be strongly correlated in frequency \citep{2019MNRAS.489..548P}. This assumption can be tested by changing the frequency of the CW source, or even injecting a comb of multiple CW signals across the band. Our expectation is that this assumption should be good enough to significantly suppress the correlated gain fluctuations on intermediate timescales however. The calibration loads will still be switched in periodically to re-measure the bandpass and re-calibrate the CW source amplitude, but this is intended to be done much less often than in systems that do not have CW sources.

\subsection{Filters and low-noise amplifiers}

RHINO targets a smaller fractional bandwidth than many other 21cm global signal experiments, in part due to the limitations of the basic horn design, and in part due to the anticipated RFI environment. The design band is 60 -- 85~MHz (34\% fractional bandwidth), with a steep rolloff required at the top of the band to suppress bright FM radio transmission that begins at 87.5~MHz. A Mini-Circuits SBP-70+ bandpass filter satisfies our requirements. In addition, a Nooelec Flamingo FM bandstop filter\footnote{\url{https://www.nooelec.com/store/downloads/dl/file/id/95/product/291/flamingo_fm_filter_datasheet_revision_1.pdf}} is included, which has an attenuation of $-30$~dB or better across the FM band.

We use Mini-Circuits ZKL-2+ low-noise amplifiers\footnote{\url{https://www.minicircuits.com/pdfs/ZKL-2.pdf}} (LNA), with a typical gain of 31.1~dB, an input VSWR of 1.02:1, an output VSWR of 1.19:1, and a noise figure of 3.4~dB, all at a reference frequency of 70~MHz. Two of these are chained together with 3~dB attenuators attached to the input and output in order to reduce reflections (the first one uses the power combiner to provide this level of attenuation at the input). This model of LNA is packaged in a metal case as standard, providing excellent shielding. We attach this to an aluminium plate to help stabilise its temperature, but do not directly control the LNA temperature beyond heating the inside of the receiver box to around $40^\circ$C to isolate it from fluctuations in the outside temperature. 

The SDR/ADC board also has its own gain of around 60~dB, with a noise figure around 2~dB (see below). A cascade analysis of the components between the switches and up to (and including) the SDR board, using datasheet values, yields an estimated noise figure of 8~dB for a maximum total gain of 112~dB. For an ambient temperature of 40$^\circ$C, this gives a noise temperature of 1660~K, which is comparable to the expected antenna temperature across the target band.



\subsection{Integrated vector network analyser}

A vector network analyser (VNA) is integrated into the system so that reflection coefficients along several signal paths can be measured in situ, rather than relying on lab measurements that occur under different environmental conditions. At the moment, this is a NanoVNA-H4\footnote{\url{https://nanovna.com/?page_id=21}} with a USB control interface. The VNA is connected to a set of SOLT (short-open-load-through) calibration targets via the main 8-way switch and an associated 2-way switch so that it can be periodically re-calibrated in-situ.

Depending on the settings of the switches, the VNA can be used to probe reflection coefficients along paths to the receiver, the antenna, and the calibration loads. {\edit An important consideration is where the reference plane of the VNA is defined. Laboratory measurements of the switches and cables may be needed to permit the VNA measurements to be transferred to a different reference plane \citep[e.g.][]{2025ExA....59....7R}. In turn, this may require enhanced measures to stabilise the switch and cable properties \citep[e.g. see][]{Zerafa} to prevent the transfer coefficients from shifting undetected.}

We note that the 8-way switch currently used by the prototype is an Analog Devices HMC253 MMIC.\footnote{\url{https://www.analog.com/en/products/hmc253.html}} While this has a reported typical port isolation of only around 35~dB at $\sim$1~GHz, the isolation is much better at low frequency. 
A high-quality mechanical switch can achieve even better isolation however, and is one of the main upgrades to be made in going from the prototype to the full receiver system.

\subsection{RF System-on-a-Chip/ADC board}


Digitisation is performed using a LimeSDR Mini 2.0 board, which combines an ECP5 FPGA and LMS7002M RF transceiver\footnote{\url{https://limemicro.com/technology/lms7002m/}} to process and digitise the input analogue RF signal. The board itself has a maximum of around 60~dB of analogue gain (adjustable), along with a 40~MHz bandwidth and maximum 12-bit sample depth. It has a USB~3.0 connection for both power and data transfer, which we connect to the backend PC via a HS02 USB isolator to suppress ground loops. The isolator is compatible with USB~2.0 `high-speed' data rates of up to 480~Mbps. The backend PC is a low-power, passively-cooled Odroid N2+ mini-PC with a quad-core Cortex-A73 CPU running Linux.

\subsection{Control, monitoring, and power systems}

The control system is based on an Arduino Due board, which drives the switches, programmes the synthesiser board, and monitors the thermistors and log-power meter. The Arduino board has a USB interface to the backend PC, and is contained within a small shielded box within the receiver box.

\section{Calibration scheme} \label{sec:cal}

In this section, we outline a calibration scheme based on a continuous wave (CW) source instead of the more usual Dicke switching approach. 

\subsection{Continuous wave calibration scheme}
\label{sec:cwcal}

To simplify the analysis, we first lump the analogue signal chain shown in Fig.~\ref{fig:sysdiagram} into a handful of generic components connected to a common reference point:
\begin{itemize}[leftmargin=0.3cm]
  \item The `receiver', defined as the part after the combiner containing the filters and LNAs; 
  \item The `antenna', defined as the antenna itself, plus the connecting cable and the relevant path through the RF switches and combiner;
  \item The `CW source', defined as the relevant path through the combiner, back to the splitter and signal generator; and 
  \item The `calibration loads', defined as the relevant path through the combiner and switches back to the heated terminations.
\end{itemize}

Each of these lumped components has a complex reflection coefficient that is a function of frequency and time. These are derived from the complex impedance $Z$ of the component with respect to a reference impedance $Z_0 = 50~\Omega$, e.g. for the receiver,
\be
\Gamma_{\rm rec}(\nu, t) = \frac{Z_{\rm rec}(\nu, t) - Z_0}{Z_{\rm rec}(\nu, t) + Z_0}.
\ee
The intention is for the relevant reflection coefficients to be measured in situ by periodically switching in the built-in VNA \citep[e.g.][]{2022NatAs...6..984D}.

Next, we can write an expression for the power measured at the output of the analogue signal chain (i.e. the output of the `receiver') when the antenna is switched in,
\be
P_{\rm rec}(\nu, t) = g(\nu, t) \big ( T_{\rm ant}(\nu, t) + T_{\rm nwave}(\nu, t) + T_{\rm cw}(\nu, t) \big ) + T_{\rm n}. \label{eq:Prec}
\ee
Here, $g$ is the gain due to the filters and amplifiers within the receiver, $T_{\rm n}$ is the noise added by the receiver, and the antenna temperature and noise wave components are given by \citep{1978ITMTT..26...34M, 2017ApJ...835...49M, 2024Univ...10..236S}
\bea
T_{\rm ant} &=& T_{\rm sky} |F_{\rm ant}|^2 \left ( 1 - |\Gamma_{\rm ant}|^2 \right ); \label{eq:Tant} \\
T_{\rm nwave} &=& T_{\rm unc} |F_{\rm ant}|^2 |\Gamma_{\rm ant}|^2 \nonumber\\
 && + \big ( T_{\rm cos} \cos(\phi_{\rm ant}) + T_{\rm sin} \sin(\phi_{\rm ant})\big ) |F_{\rm ant}| |\Gamma_{\rm ant}|, \label{eq:Tnwave}
\eea
where we have now suppressed the arguments for brevity. $T_{\rm sky}$ is the true sky temperature following convolution with the instrumental antenna pattern. The $T_{\rm unc}$, $T_{\rm cos}$, and $T_{\rm sin}$ factors are noise wave parameters that account for noise from the receiver that has been reflected back off the antenna and into the receiver again. They are, respectively, the uncorrelated, and real/imaginary parts of the correlated components of this noise signal (where we mean correlated or uncorrelated with the output of the receiver that would be obtained if there were no reflections). All of these factors have units of Kelvin, and are functions of time and frequency that must be determined as part of the calibration scheme. We have also defined
\be
F_{\rm ant} = \frac{\sqrt{1 - |\Gamma_{\rm rec}|^2}}{1 - \Gamma_{\rm ant} \Gamma_{\rm rec}}.
\ee
For the CW source term, $T_{\rm cw}$, we add in the power that it contributes, plus the reflected noise wave signals. These will have a different reflection coefficient, $\Gamma_{\rm cw}$, as the signal path to the CW source is different from the antenna one. In the first instance, we neglect multi-path reflections, e.g. signals from the CW that have been reflected off the receiver, then back to the antenna, and then back to the receiver. Because the CW source can have a high effective temperature, these may need to be included however. The CW signal is output through a splitter, and then a combiner, each of which contributes around 3~dB of attenuation each time a signal passes through (in either direction). We therefore expect the impedance match to be relatively good. In this case we can write $\Gamma_{\rm cw} \approx 0$, so the CW source term (including reflections) is
\be
T_{\rm cw} \approx T_{\rm cwsrc} \big ( 1 - |\Gamma_{\rm rec}|^2 \big ),
\ee
where $T_{\rm cwsrc}$ is an effective temperature of the CW source, i.e. the signal generator tone after losses through the splitter, combiner, and connecting coaxial cables.

The purpose of the CW source is to provide continuous monitoring of the time dependence of the gain, $g$, in a single frequency channel, while the calibration sources provide an absolute power level reference and a bandpass calibration. The CW source is monitored by a logarithmic power meter attached to the splitter, which measures power integrated over a wide band. We write this power as
\be
P_{\rm lpm} \approx g_{\rm lpm} (T_{\rm cwsrc} + T_{\rm base}),
\ee
where $g_{\rm lpm}$ is a gain factor for the log power meter that accounts for losses due to the splitter etc., and $T_{\rm base}$ is the baseline power input into the meter when the signal generator is switched off.
The power meter is assumed to be well-calibrated and highly stable, so that any instability occurs on significantly longer timescales than the knee frequency of the correlated ($1/f$) receiver noise that we are trying to calibrate out. The gain factor $g_{\rm lpm}$ and baseline power are also assumed to be stable on similarly long timescales. By subtracting the baseline power, $P_{\rm base} = g_{\rm lpm} T_{\rm base}$, measured while the signal generator is switched off (at $t_{\rm off}$), and dividing through by the power meter's output at a reference time $t_0$, we obtain
\be
y(t) \equiv \frac{P_{\rm lpm}(t) - P_{\rm base}(t_{\rm off})}{P_{\rm lpm}(t_0)  - P_{\rm base}(t_{\rm off})} =  \frac{g_{\rm lpm}(t)}{g_{\rm lpm}(t_0)} \frac{T_{\rm cwsrc}(t)}{T_{\rm cwsrc}(t_0)} \approx \frac{T_{\rm cwsrc}(t)}{T_{\rm cwsrc}(t_0)}.
\ee
This quantity tracks the time dependence of the CW signal. The splitter also injects the CW signal into a single frequency channel of the receiver (the $T_{\rm cw}$ term in Eq.~\ref{eq:Prec}). We can isolate this contribution by differencing neighbouring frequency channels in the receiver output to obtain
\bea
\delta P_{\rm cw}(t) &=& P_{\rm rec}(\nu_{\rm cw}, t) - P_{\rm rec}(\nu_{\rm cw} + \delta\nu, t) \\
 &\approx& g(\nu_{\rm cw}, t)\, T_{\rm cwsrc} \big ( 1 - |\Gamma_{\rm rec}|^2 \big ) \\
 & \propto & g(\nu_{\rm cw}, t)\, y(t).
\eea
Hence, we can estimate an overall time-dependent gain factor as
\be
\frac{\delta P_{\rm cw}(t)}{\delta P_{\rm cw}(t_0)} = \frac{g(\nu_{\rm cw}, t)}{g(\nu_{\rm cw}, t_0)} y(t).
\ee
This is independent of frequency. Note that schemes based on differencing can be biased by noise fluctuations and smooth (but non-zero) variations in the continuum power between neighbouring channels. Instead of directly differencing with neighbouring channels, one can instead fit a continuum across multiple channels and de-trend. Using multiple channels also helps average down effects due to noise. Beyond this, a full forward-modelling approach to the calibration permits self-consistent estimates of these effects to be made \citep[c.f.][]{2021MNRAS.505.2638R}. The real-world performance of the RHINO CW calibration, and different algorithmic approaches to measuring correlated gain fluctuations, will be examined in a future work.

The CW approach can track temporal variations in the gains that are strongly correlated across the band, but would become ineffective if there is substantial spectral variation in the $1/f$ noise. Related issues were examined by \cite{2019MNRAS.489..548P}, and solutions such as injecting a frequency comb (instead of a single tone) can be explored if necessary. Establishing whether the gain fluctuations are indeed strongly correlated in frequency is a key priority for the RHINO prototype phase.


\subsection{Warm calibration loads}
\label{sec:loadcal}

A bandpass calibration is also needed, and cannot be provided by the CW source. For this, we use the more traditional approach of periodically observing a termination heated to a known temperature, i.e. a warm or hot calibration load. This will produce a total noise power of $P_{\rm load} = k_{\rm B} T_{\rm load} \Delta \nu$ over a bandwidth $\Delta \nu$. Because we do not need to rapidly switch between the calibration load and the antenna, we can switch to the load periodically, dwell on it for longer, and get a lower-noise measurement of its reference power. If the effective temperature of the load is well-measured, and any additional spectral shape due to reflections is known, this gives us a bandpass calibration. With two loads of different temperatures to switch between, we can derive an absolute temperature calibration.

We can write similar expressions as the above for the calibration loads. Since these are simply 50~$\Omega$ terminations, we also expect their match to be quite good, such that $\Gamma_{\rm cal} \approx 0$. If this approximation applies, we can then write
\be
T_{\rm cal} \approx T_{\rm calsrc} \big ( 1 - |\Gamma_{\rm rec}|^2 \big ),
\ee
where $T_{\rm calsrc}$ is the effective temperature of the termination. This is derived from the physical temperature of the termination plus a correction due to the temperature of the (lossy) cable that connects it to the receiver. For a linear temperature gradient between the calibration source and the receiver's input port, and a cable insertion loss below 0.5~dB, we can make the approximation \citep{1966ARA&A...4...77F},
\be
T_{\rm calsrc} \approx T_{\rm calsrc}^{\rm phys} + \Delta T \left ( 0.1152 |L| - 0.0088 |L|^2 \right ),
\ee
where $\Delta T = T_{\rm rec}^{\rm phys} - T_{\rm calsrc}^{\rm phys}$ is the difference between the physical temperatures of the lossy cable at the receiver input port and the calibration source, and $L$ is the insertion loss of the cable in dB. By using short lengths of semi-rigid coaxial cable with SMA connectors, we hope to minimise the insertion loss to below $L \lesssim 0.3$~dB. For a calibration load at 383~K, receiver port at 303~K, and insertion loss of 0.3~dB, the difference between the physical and effective temperature of the calibration source would then be $\approx -2.7$~K, or about a 0.7\% correction to the calibration source power level. Stabilising the temperature of the cables/receiver, as well as the calibration source, is therefore necessary to avoid fluctuations in the reference power level at the sub-percent level. Note that the effect of the cables also includes reflections due to inexact matching, as well as signal attenuation and added noise power \citep{2008ApJ...688...12Z}.

The choice of calibration load temperatures is discussed in Sect.~\ref{sec:frac_obs}. A substantially higher temperature is preferable for the hottest load. This presents a number of challenges in terms of temperature control, safety, and the materials used however; the PTFE dielectric commonly used in coaxial cable has a melting point of around 600~K for example. Instead, it may be more practical to use a non-thermal reference source instead, such as a noise diode. These can achieve much larger effective temperatures, with good temporal stability, but would need an additional bandpass calibration of the source to be performed. 

\subsection{Optimisation of observing time/calibration cycles} \label{sec:frac_obs}
In this section, we discuss an optimised way of choosing how much time should be spent observing each calibration load vs the antenna.

Pairs of well-characterised reference sources can be used to calibrate a given source by forming a quotient
\be
\label{eq:PSD_Quotient}
    q =  \frac{ P_{\text{src}} - P_{\text{L}}}{P_{\text{NS}} - P_{\text{L}}},
\ee
where $P_{\text{src}}$ is a measurement of the power spectral density (PSD) for a source input to the receiver, and $P_{\text{NS}}$ and $P_{\text{L}}$ are PSDs from measurements of two different reference noise-sources of known noise temperature. 
Such quotients can then be used to fit gains, noise-wave parameters, and other degrees of freedom needed to calibrate the noise temperatures of an input source such as an antenna \citep{2012RaSc...47.0K06R, 2021MNRAS.505.2638R, 2022MNRAS.517.2264M}.

Eq.~\ref{eq:PSD_Quotient} is a first-order approximation that assumes that the two reference noise sources are perfectly impedance-matched to the receiver. This removes dependence on signal gains that occur after the switching interface, as well as the self-generated noise power from the receiver that does not undergo reflections before being measured.
Additionally, the assumption is made that the PSD measurements are taken within a time frame in which the receiver gains are constant. This is equivalent to keeping the total integration time under the knee-frequency of the receiver so that the PSD measurements are dominated by Gaussian thermal noise.

The signal-to-noise ratio (SNR) when measuring $q$ limits the precision to which the receiver noise-wave parameters can be measured, which in turn affects the calibrated value of the sky temperature, $T_{\text{sky}}$. It is therefore beneficial to maximise this SNR if possible. This can be achieved by tuning the integration times for the PSD measurements as well as the choice of calibrator noise temperatures.

The optimal observing strategy is well-known in Dicke-switched systems, for example \citep{dicke1946measurement}. Observing time is split equally between the source and calibrator, and when $T_{\rm src} \approx T_{\rm L}$, it can be shown that the effects of gain fluctuations are cancelled out when observing a source and calibrator of the same radiometric power.

We now present a derivation of an optimised calibration strategy for our receiver system, subject to the assumptions above. Assuming that the three PSD measurements for a single estimate of $q$ are completed in a total time $\tau$, with a channel bandwidth $\Delta\nu$, the radiometer equation can be used to derive the variance for each measurement based on the total system noise temperature. The variance of a single frequency channel for a PSD measurement of a given source, $i$, can be expressed as

{\edit
\be
\sigma_{P_i}^2 = \left (g k_{\rm B} T_{\text{sys}}\right )^2\frac{\Delta\nu}{\phi_i\tau} = \frac{P_i^2}{\Delta \nu \phi_i \tau},
\ee

}
where $g$ is the receiver gain, $k_{\rm B}$ is Boltzmann's constant, $T_{\text{sys}}$ is the receiver system temperature and $\phi_i$ is the fraction of the total time $\tau$ spent on the single PSD measurement such that $\phi_{\text{src}}+\phi_{\text{L}}+\phi_{\text{NS}}=1$.

Neglecting gain fluctuations and spectral leakage between channels in this analysis (which introduce correlations in frequency and time), when the error on the PSDs are propagated in quadrature, the fractional variance for a measurement of $q$ can be expressed as

{\edit
\bea
\label{eq:q_fractional_variance}
\left( \frac{\sigma_q}{q} \right)^2 = \frac{1}{\Delta\nu\tau} \left[ \frac{P_{\rm src}^2}{(P_{src} - P_{\rm L})^2} \frac{1}{\phi_{\rm src}} \right.  
+ \frac{P_{\rm NS}^2}{(P_{\rm NS} - P_{\rm L})^2} \frac{1}{\phi_{\rm NS}} \nonumber\\ 
+ \left. \frac{P_{\rm L}^2 \left(P_{\rm src} - P_{\rm NS}\right)^2}{\left( P_{\rm NS} - P_{\rm L} \right)^2 \left( P_{\rm src} - P_{\rm L} \right)^2 } \frac{1}{\phi_{\rm L}} \right],
\eea
where $P_{\rm src}$, $P_{\rm NS}$ and $P_{\rm L}$ retain the definitions defined in Eq. {\ref{eq:PSD_Quotient}}. This can be expressed in terms of noise temperatures and reflection coefficients for the receiver and source by substituting for Eq. {\ref{eq:Prec}}.
}
\begin{figure}
    \centering
    \includegraphics[width=1\columnwidth]{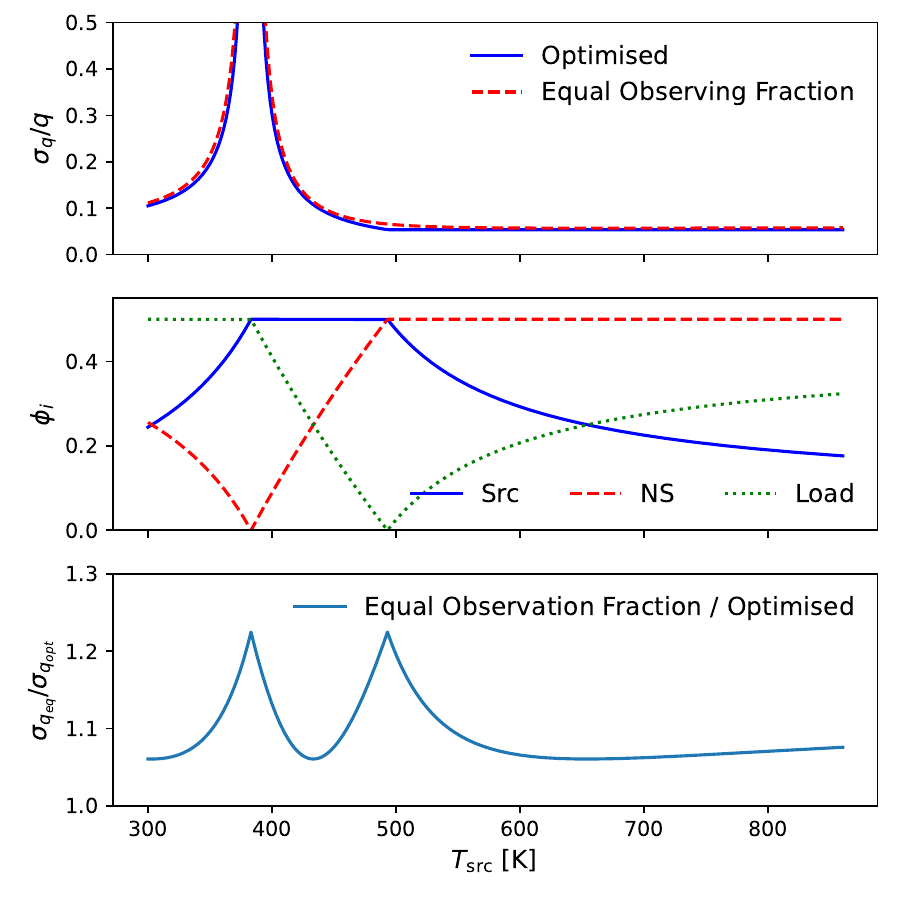}
    \caption{The predicted fractional standard deviation for a single measurement of $q$ when $T_{\rm L}=$ 380~K, $T_{\rm NS}=$ 490~K, the total observing time $\tau=$45~s and $\delta\nu=$1.22~kHz. The bottom plot shows the ratio between the standard deviations of $q$ for the optimised equal strategy and the optimised observation strategy.}
    \label{fig:frac_obs}
\end{figure}

Eq.~\ref{eq:q_fractional_variance} can be minimised with respect to the fractional observation times $\phi_i$ subject to the constraint $\phi_{\text{src}}+\phi_{\text{L}}+\phi_{\text{NS}}=1$, to yield

{\edit
\be
\hat{\phi_i} = \alpha m_i c_i
\ee
where 
\bea
m_{\rm src} &=& \frac{P_{\rm src}}{P_{\rm src} - P_{\rm L}},\\
m_{\rm NS} &=& \frac{P_{\rm NS}}{P_{\rm NS} -  P_{\rm L}}, \\
m_{\rm L} &=& \frac{P_{\rm L} \left( P_{\rm src} - P_{\rm NS} \right) }{\left( P_{\rm NS} - P_{\rm L} \right) \left( P_{\rm src} - P_{\rm L} \right) }, \\
\alpha &=& \left( m_{\rm src} + c_{\rm NS}m_{\rm NS} + c_{\rm L}m_{\rm L} \right)^{-1}
\eea
and
\[
c_{\rm src}, c_{\rm NS},c_{\rm L} = 
\begin{cases}
    +1, -1, -1 & \text{if } P_{\rm src} \leq P_{\rm L} \\
    +1, +1, -1 & \text{if } P_{\rm L} < P_{\rm src} \leq P_{\rm NS} \\
    +1, +1, +1 & \text{if } P_{\rm NS} < P_{\rm src}.
\end{cases}
\]
}

Fig.~\ref{fig:frac_obs} shows a comparison between the optimal fractional observation times and the case in which equal time is spent on each target ($\phi_{\text{src}}=\phi_{\text{L}}=\phi_{\text{NS}}$) as a function of $T_{\rm src}$. {\edit Here, $P_{\rm L}$ and $P_{\rm NS}$ take values sourced from the heated impedance-matched loads outlined in Sect.~\ref{sec:WarmHotLoads} ($T_{\rm L} = $ 380~K and $T_{\rm NS} = $ 490~K). The channel bandwidth, $\delta\nu$, is set to be 1.22~kHz, the total observing time is set to $\tau=$ 45\,s and the gain is set to unity. $P_{\rm src}$ is calculated assuming that $P_{\rm src} = gk_B\Delta\nu (T_{\rm src}+T_0)$ where $T_0$ is the receiver-induced, non-reflected noise temperature (reflection terms are absorbed into $T_{\rm src}$).} 

In both cases there are singularities when $T_{\rm src} = T_{\rm L}$, resulting in a sharp increase in the variance of $q$ and a resulting decrease in the SNR. Although the SNR can be increased with longer integration times, in actual observations this is limited to the time scale on which the gain fluctuations dominate over the thermal white noise ($\tau \lesssim  1/f_{\rm knee}$).

Optimising the fractional observation times when $T_{\rm src} \approx T_{\rm L}$ can lead to a significant improvement in the SNR on the quotient, of a little over 20\% compared with the equal-time case. When $T_{\rm src}$ becomes much larger than $T_{\rm NS}$, the improvement over the equal-time case is only around $7-8$\% however. This shows a relative insensitivity to the fraction of time spent observing each calibrator, implying that longer can be spent observing through the antenna without significantly degrading the calibration measurements. 

As $T_{\rm src}$ may also encapsulate reflected noise power contributions, these must also be considered when attempting to improve the SNR for measurements of $q$. Therefore, an approximation of the expected noise temperature of $T_{\rm src}$, including the reflected noise wave terms using Eqs. \ref{eq:Tant} and \ref{eq:Tnwave}, can be used to ensure a suitable difference in noise temperature from impedance-matched calibrator $T_{\rm L}$. As this depends on measurements of the reflection coefficients and receiver noise wave parameters, this tuning of $T_{\rm L}$ may follow from laboratory measurements of receiver and source properties.

To extract the noise wave parameters of the receiver, measurements of $q$ can be made for calibration standards such as long open and shorted cables, as well as other impedance mismatched calibrators, to induce known reflections and phase delays, producing spectral features that can be measured and used to derive a calibration \citep{2021MNRAS.505.2638R, 2023arXiv230511479P}. These measurements are typically made for standards at an ambient temperature $T_{\rm amb}$, with a resulting noise temperature $T_{\rm src} \approx T_{\rm amb}$. Tuning the noise temperature of $T_{\rm L}$ to be sufficiently higher than $T_{\rm amb}$ (so that the total $T_{\rm src}$, including the reflected noise wave terms, does not exceed or approach $T_{\rm L}$ for multiple calibration standards) will result in a decrease in noise on the measurements of $q$. This in turn will result in an increase in accuracy of measurements of the induced spectral features and subsequent extraction of the noise wave parameters through fits and other methods. The calibrated $T_{\rm sky}$ will then be more precise, with less contamination from the reflected receiver noise power.

In the fully optimised case, in which the highest SNR is extracted with maximal time spent observing the sky, $T_{\rm NS}$ should exceed the expected antenna noise temperature. Given the typical sky temperatures of $\gtrsim 10^3$~K at these frequencies, this would need to be achieved with a non-thermal source, e.g. a noise diode.

\subsection{Calibration requirements} \label{sec:calspecs}

In general, the precision with which we can determine the noise wave calibration parameters will be a function of temporal and frequency scale. There is a relatively complex interaction between calibration and the foreground removal and 21cm model fitting procedures, as calibration errors can modulate these signals, producing spurious features that can bias or be confused with 21cm absorption features for example. Additive systematics can also partially mimic or obscure the cosmological signal. The way in which these interactions play out depends on the methods used to propagate the calibration, remove the foregrounds, and extract the 21cm signal -- for instance, data-driven methods reliant on spectral smoothness may absorb the effects of calibration errors differently from more constrained methods based on forward modelling. As such, it is difficult to put strict calibration requirements on the system a priori, as they will depend on a variety of analysis choices that we will need to make further down the line, i.e. once the system is operational and we have a good idea of the behaviour of the receiver. Other groups have performed simulated analyses \citep[e.g.][]{2022NatAs...6..984D}, subject to this caveat.

\begin{figure}
    \centering
    \includegraphics[width=\columnwidth]{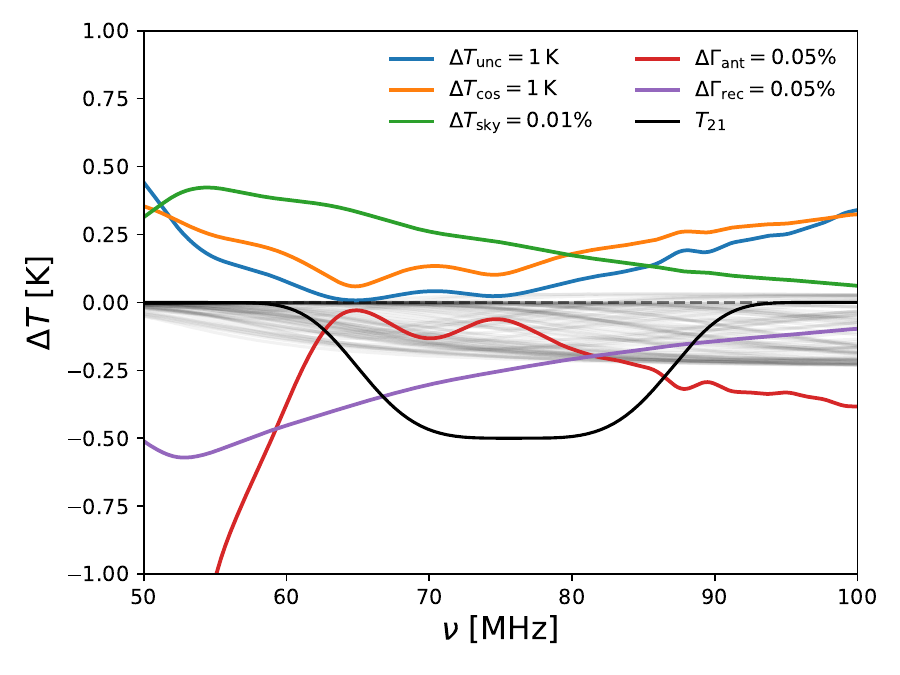}
    \caption{\edit Temperature difference vs frequency caused by mis-estimates of various noise wave parameters. The values have been chosen to give a temperature difference of similar size to {\edit the physical 21cm models described in Sect.~\ref{sec:basicsims} and Fig.~\ref{fig:theory}}, shown as faint grey lines. An EDGES-like absorption feature is also shown as a solid black line.}
    \label{fig:dT_noisewave}
\end{figure}

Nevertheless, we have tried to give some sense {\edit of the importance} of errors on different calibration parameters, and how these could obscure the 21cm signal. Fig.~\ref{fig:dT_noisewave} shows the 21cm global signal for a set of example physical models, plotted with various systematic and calibration error terms. The amplitude of these terms is chosen to indicate the approximate measurement precision required to restrict each term to a typical value of around 100~mK, i.e. comparable to the expected global signal. Depending on the real-world spectral and temporal characteristics of each of these terms, they may be more or less difficult to separate from the 21cm signal, so these numbers should be taken only as a guide, but we can at least get a sense of the size of calibration errors that could be tolerated without severely affecting the 21cm signal extraction. Note that we have omitted several important potential sources of calibration error, including the CW source and the effects of soil conductivity/groundspill, as these will be the subject of much more detailed treatments in forthcoming publications.

\begin{figure*}
    \centering
    \includegraphics[width=2\columnwidth]{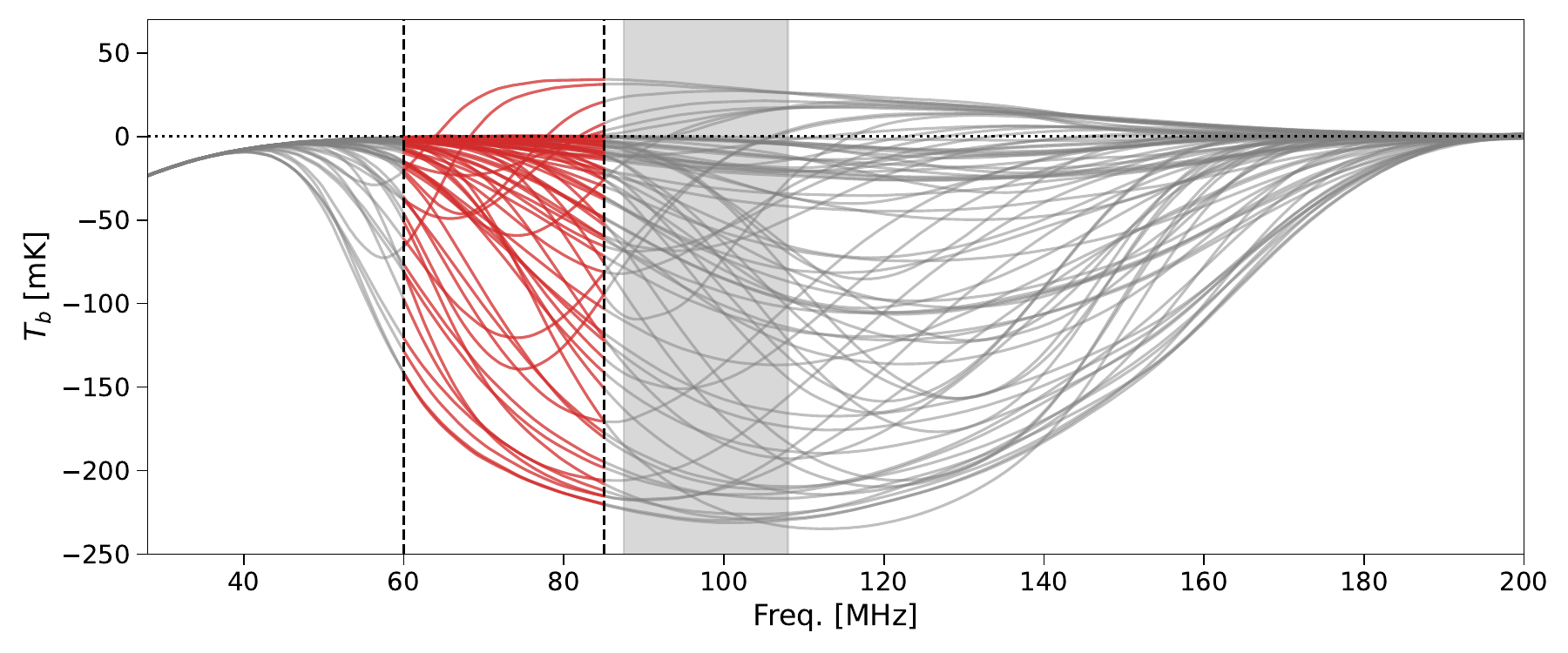}
    \caption{Predictions of the 21cm global signal for 80 randomly-selected models from the 21cmVAE emulator (grey), with the frequency window that RHINO is sensitive to highlighted in red. The grey region shows the FM band, which is the most serious contaminant, although other RFI sources are also present at higher and lower frequencies.}
    \label{fig:theory}
\end{figure*}

To derive these illustrative examples we use Eqs.~\ref{eq:Tant} and \ref{eq:Tnwave}, with EDGES-like values for the noise-wave parameters \citep[see Extended Data Figure 5 of][]{2018Natur.555...67B}, and the antenna reflection coefficient derived from the calculated $S_{11}$ shown in Fig.~\ref{fig:s11}. All of these parameters will of course differ when the system is deployed, but these choices should be sufficiently realistic for this illustration. We then introduced a small error on each parameter in turn, setting all other parameters to their fiducial values, and then subtracted off the fiducial model to leave a temperature difference $\Delta T$ due to each term. The size of the error was chosen by hand to make the typical value of $\Delta T \sim 100$~mK, and is shown for each term in the legend of Fig.~\ref{fig:dT_noisewave}. For further comparison, we also plotted a simple illustrative error due to an imperfect foreground model, represented by a systematic fractional error of $0.01\%$ on the temperature of each pixel of the sky model at each frequency.

It can be seen from Fig.~\ref{fig:dT_noisewave} that the calibration requirements are quite stringent, and that the calibration errors (in this simple approach) are spectrally smooth, and so do risk introducing spectral features that could be at least partially confused with the 21cm global signal. The noise wave temperature coefficients would need to be constrained to 0.1~K or better, while the antenna and receiver reflection coefficients must be measured to better than 0.01\% (errors below $-40$~dB). The latter should ultimately be achievable given that the dynamic range of modern VNAs exceeds this level by tens of dB. In comparison, the simple foreground model error shown here highlights the difficulty of building sufficiently precise forward models of the sky; a mere 0.01\% error is enough to produce a temperature difference of similar size to the expected 21cm signal. Notably, the `typical' expected 21cm signal is smooth and slowly-varying across our band, increasing the difficulty of distinguishing it from calibration errors compared with the more distinctive EDGES signal.

\section{Performance forecasts} \label{sec:forecasts}

In this section, we show a set of simple forecasts for RHINO's ability to recover physical 21cm global signal model parameters. We take an idealised approach, simulating only diffuse foregrounds and Gaussian thermal noise, plus a 21cm global signal model drawn from an emulator. We then perform a parameter inference study, using a Markov Chain Monte Carlo (MCMC) method to jointly fit a foreground and 21cm signal model to the time-averaged simulated data. Imperfections such as calibration errors and RFI flagging are not included in the simulations at this stage, and will be the subject of a more detailed future work. We compare the recovery of parameters from scenarios where no foregrounds are present in the data, and where diffuse foregrounds are present, and are removed using a `blind' PCA-based removal method.

\begin{figure*}
    \centering
    \includegraphics[width=2\columnwidth]{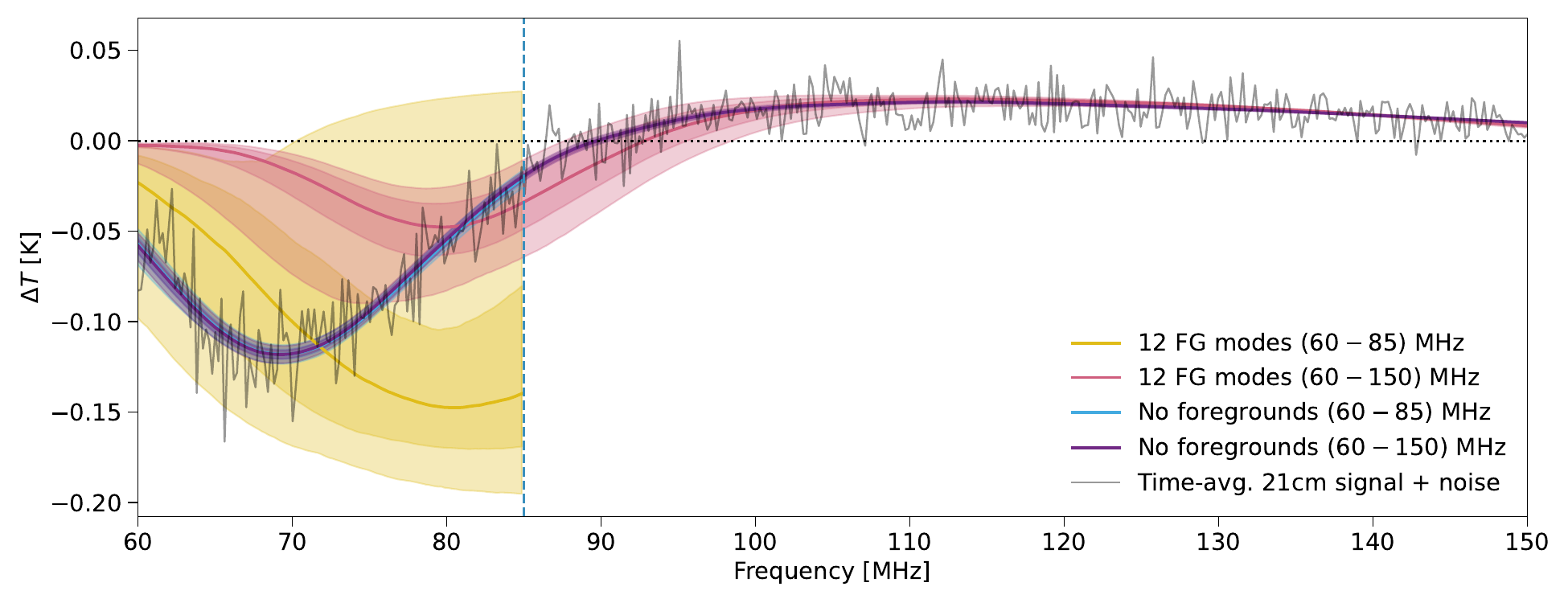}
    \caption{Posterior predictive distributions for the global signal as predicted by {\tt 21cmVAE}, for the RHINO ($60 - 85$~MHz) and wideband ($60-150$~MHz) scenarios, for simulated data with perfect calibration, no RFI flags, and thermal noise equivalent to 20 full nights of observing. To permit a fair comparison across the full frequency range, a frequency-dependent Airy beam for a 7m aperture was assumed. Fits are shown to simulated data without foregrounds, and with PCA-subtracted foregrounds (12 modes removed). The median is shown as a solid coloured line, and the 68\% and 95\% credible regions are shown as shaded bands. The true (input) global signal is shown as a solid grey line with time-averaged noise included.}
    \label{fig:modelfits}
\end{figure*}

\subsection{Basic simulations and model fitting} \label{sec:basicsims}

Compared to other 21cm global signal experiments, RHINO has a relatively narrow design bandwidth of 25~MHz. This is partly limited by the physical size of the horn at the low frequency end, and the bright FM radio band at the high frequency end. Physical models of the 21cm global signal predict absorption features that can be substantially wider than this however; widths of order tens  of MHz are typical, although narrower features are also possible.

Our focus here is on recovery of physical model parameters for `standard' models that do not attempt to explain the EDGES signal. We use the {\tt 21cmVAE} emulator \citep{2022ApJ...930...79B} to define a broad space of possible global signal models, parametrised by 7 values that summarise pertinent physical processes, such as the star formation efficiency, $f_*$, X-ray heating efficiency, $f_{\rm X}$, the minimum circular velocity for a halo to support star formation, $V_c$, and the low-energy cut-off of the X-ray SED, $\nu_{\rm min}$ \citep{2020MNRAS.495.4845C}. Most of the variation in the global signal can be captured by these four parameters; the other 3 parameters (ionising photon mean free path, $R_{\rm mfp}$, optical depth to the CMB, $\tau$, and X-ray SED power-law slope, $\alpha$) have a relatively minor effect on the global signal \citep{2022ApJ...930...79B}.

Fig.~\ref{fig:theory} shows 80 randomly-chosen examples of plausible models produced by the emulator. The parameter values were randomly drawn from uniform distributions with ranges at or close to the maximum ranges of the {\tt 21cmVAE} emulator, i.e. $f_* \in [10^{-4}, 0.5]$, $V_c \in [4.2, 100]$~km/s, $f_X \in [10^{-5}, 100]$, and $\nu_{\rm min} \in [0.1, 3.0]$~keV, with the other parameters fixed to values of $\tau = 0.07$, $\alpha=1.25$, and $R_{\rm mfp} = 29.6$~Mpc, as in \cite{2022ApJ...930...79B}. The RHINO band is highlighted in red, and while it can be seen that some models have narrower absorption features that fall within this region, many others have broader features, leaving only a relatively smooth and featureless gradient across the RHINO band. The smoother the global signal is, the harder it is likely to be to disentangle it from similarly smooth foregrounds and systematic artefacts. Note that the emulator predictions for the global signal are noisy at small frequency separations, and so we apply a Savitzky-Golay filter \citep{1964AnaCh..36.1627S} to smooth out the small-scale variation, which would otherwise be confused with thermal noise.

Next, we chose one of the randomly generated models that has an absorption feature roughly 100~mK deep within the RHINO band (see Fig.~\ref{fig:modelfits}). This was added to basic simulations that used a frequency-dependent Airy beam pattern for a 7.2m aperture telescope, a sky model containing only diffuse emission from GSM 2016 \citep{2017MNRAS.464.3486Z}, and thermal noise consistent with 20 nights of observing with short drift scans of 72 minutes with 10s integrations. Instrumental systematics and calibration errors were not included; while these are important effects, we only seek to understand the impact of the RHINO bandwidth on simple idealised scenarios in this study.

With the simulations in hand, we performed a joint foreground plus 21cm global signal model fit. The 21cm models were provided by {\tt 21cmVAE}, with 4 free parameters as discussed above. The foregrounds were modelled using an idealised Principal Component Analysis (PCA) technique, which we implemented by taking the simulated foreground model (including beam effects, but without noise) for the relevant frequency window for each scenario and calculating the first few principal components of its frequency-frequency covariance matrix (constructed by averaging over time). The first 12 PCA modes were used to define the foreground model that is included in the fits. These modes had significant smooth spectral variation based on visual inspection, whereas subsequent modes became more noise-like. The PCA mode amplitudes and 4 global signal model parameters were then jointly sampled using {\tt emcee} \citep{2013PASP..125..306F}, with uniform priors on all parameters. For simplicity, we worked with the time-averaged data for the model fitting, although one could in principle attempt to fit the foreground separately for each time sample.

We ran {\tt emcee} with 48 workers for 2000 samples each. The chains were initialised close to the best-fit parameter values, and after visual inspection, we removed only 100 burn-in samples per chain. The convergence of the individual chains (for each worker) was quite variable, in large part due to some of the parameters being poorly constrained, but the ensemble of workers as a whole appeared to have explored the posterior thoroughly enough. Increasing the number of MCMC samples did not change the results appreciably.

\subsection{Global signal recovery} \label{sec:recovery}

The posterior predictive distribution (PPD) of the global signal model is shown in Fig.~\ref{fig:modelfits}, for an experiment with a RHINO-like 25~MHz bandwidth (yellow/blue), and another experiment with a much broader bandwidth of 90~MHz (red/purple). For foreground-free data, the recovery of the true signal (shown in grey, with time-averaged noise) is excellent across the whole frequency range, and there is little difference between the two bandwidths except slightly tighter credible regions for the broadband case.

\begin{figure}
    \centering
    \includegraphics[width=\columnwidth]{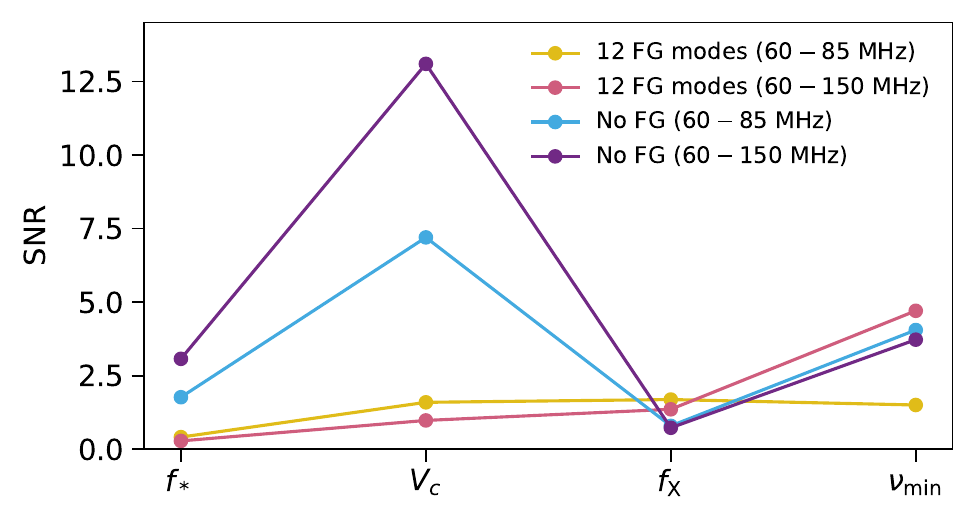}
    \caption{Marginal parameter constraints for the scenarios shown in Fig.~\ref{fig:modelfits} for the four parameters of the {\tt 21cmVAE} model that were not fixed. The constraints are expressed as an effective signal to noise ratio for each parameter, defined as the fiducial parameter value divided by the standard deviation of the marginal distribution.}
    \label{fig:paramsnr}
\end{figure}

\begin{figure*}
    \centering
    \includegraphics[height=6cm]{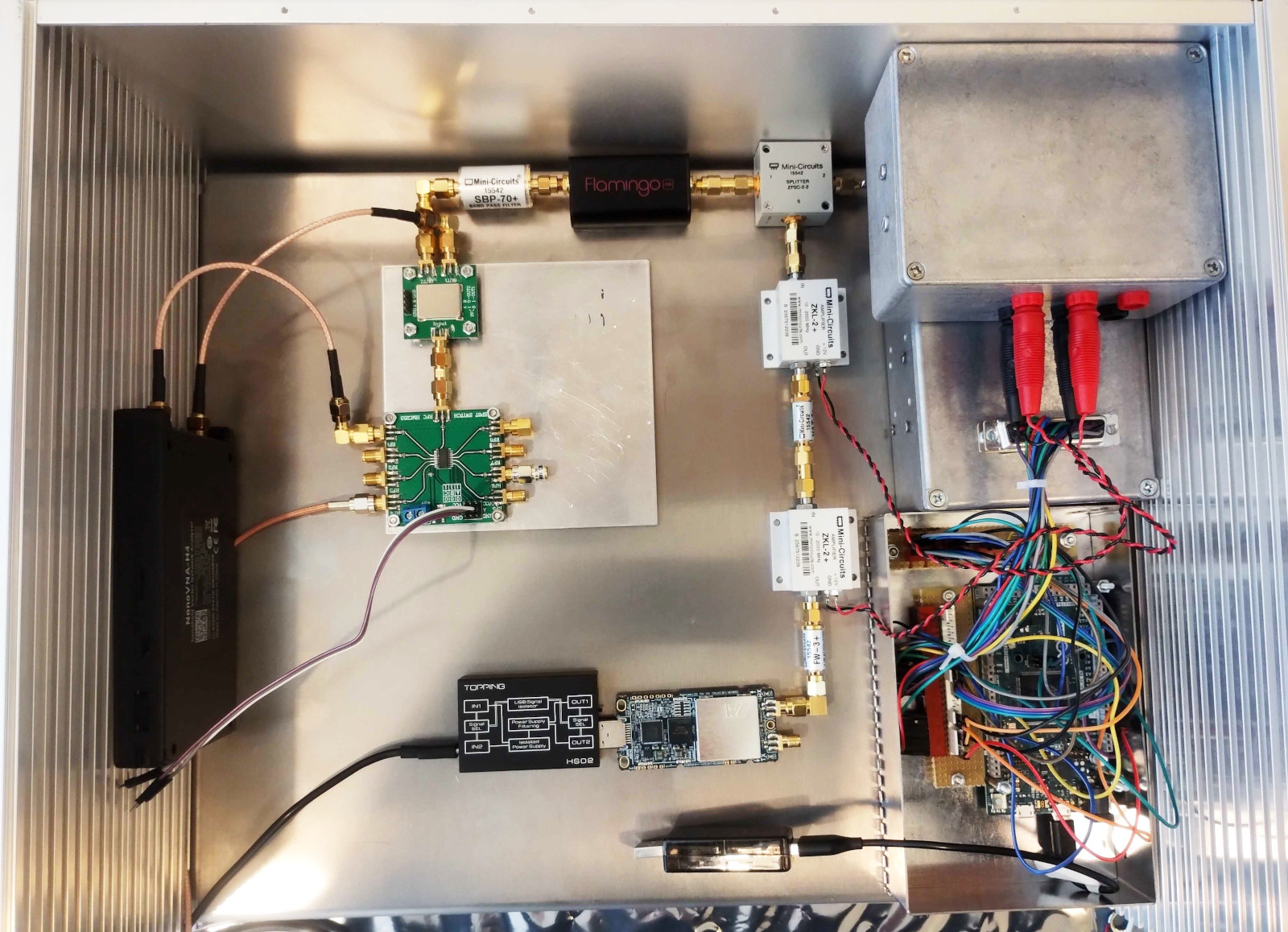}
    \includegraphics[height=6cm]{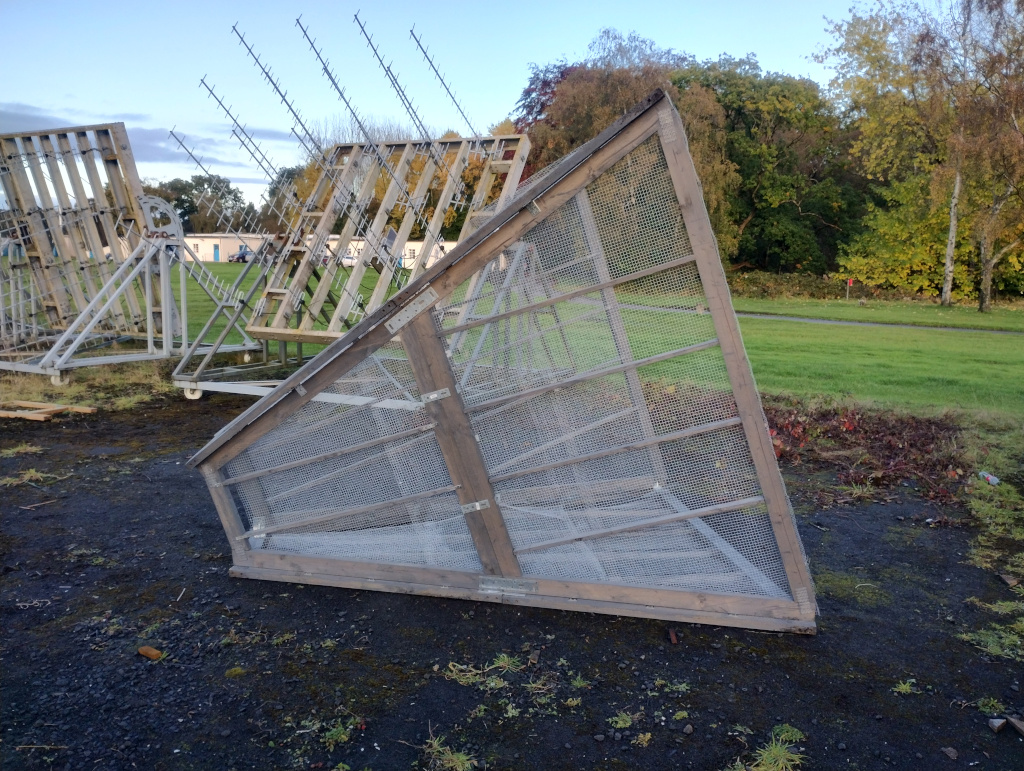}
    \caption{{\it (Left):} Prototype receiver, laid out on an aluminium plate that fits into a shielded box. The VNA is on the left wall; RF switches near the centre; RF filters and splitter at the top; LNAs centre-right; ADC board and USB isolator at the bottom; and power supply, shielded CW box, and Arduino controller (with shield lid off) along the right wall.
             {\it (Right):} Side view of the scaled-down prototype horn as constructed (flare section only).}
    \label{fig:prototype}
\end{figure*}

For the blind foreground-subtracted data, however, the credible regions are significantly expanded. In the broadband case, the credible regions are narrower, but the (model dependent) reconstructed 21cm global signal is biased at the lowest frequencies, around the absorption feature. The recovered absorption feature is shifted by more than 10~MHz, and is less than half of its true amplitude. The results are therefore biased in a consequential manner. The recovery is quite good at the higher frequencies however, and the inference recovers the long tail of emission at $\nu \gtrsim 90$~MHz very well. For the RHINO-like case, the credible regions are broader, and there is not a significant detection -- the 95\% interval includes $\Delta T = 0$~mK. The true model resides comfortably within this interval (i.e. there is not a bias as in the broadband case), although this is mostly because the RHINO case makes only a mild improvement over the prior.

Both of these cases demonstrate the need for robust physical modelling of the instrument and sky signals. Blind (data-driven) foreground removal methods permit too much flexibility in the foreground model, resulting in most of the 21cm global signal being absorbed by the foreground modes. This results in strong correlations between the foreground and global signal parameters, making it difficult to unambiguously disentangle them. Physical modelling \citep[e.g.][]{2017AJ....153...26S} should result in a much more constrained range of possible foreground behaviours, at the expense of significantly increased model complexity. From the results shown in Fig.~\ref{fig:modelfits}, we can conclude that much more constrained foreground models will be necessary if RHINO is to be able to recover the global signal -- the limited frequency range increases the confusion between foregrounds and signal. Similar measures would be needed even for an experiment with much wider bandwidth, however, as strong correlations between foreground modes and the smooth global signal persist, resulting in a biased inference in the particular example shown above. {\edit Recent work on Bayesian sky models for 21cm global signal applications offers a promising approach to address this issue \citep[e.g.][]{2025arXiv250101417C, 2025arXiv250621258I, 2025arXiv250911894N}.}

Fig.~\ref{fig:paramsnr} shows the effective signal to noise ratio on the four global signal model parameters that were included in the inference, for each of the four scenarios discussed above. The X-ray heating efficiency parameter, $f_{\rm X}$, was unconstrained in all four cases, even for idealised foreground-free data. All of the parameters were only weakly constrained at best in the blind foreground removal scenarios, except for $\nu_{\rm min}$, which was moderately well-measured in the broadband case. This suggests that there is some value to extending observations to above 100~MHz, even in the pessimistic situation where we have a poor/overly-flexible foreground model. In the highly optimistic foreground-free case, the $f_*$, $V_c$, and $\nu_{\rm min}$ parameters can all be measured, although only $V_c$ is strongly constrained. Increasing the bandwidth improves the constraints, as one would expect, although not dramatically so.

\section{Prototype system} \label{sec:prototype}

A prototype system is under construction at Jodrell Bank Observatory (JBO). This uses a scaled-down horn antenna, and targets a correspondingly higher frequency band, using essentially the same receiver design except for a different set of filters. In this section, we briefly outline the progress to date in its construction and testing.

\paragraph*{Prototype receiver:} The left panel of Fig.~\ref{fig:prototype} shows the prototype receiver laid out on an aluminium plate. This follows the same design as in Sect.~\ref{sec:system}, and as detailed above, mostly uses off-the-shelf connectorised components or printed circuit boards that we mounted in connectorised shielded boxes. 
The main purpose of this prototype is to develop and test a functioning calibration scheme, and permit observations with the scaled-down prototype horn antenna. The former may require components to be upgraded as limitations of the system are discovered. The latter will necessitate a different choice of filters to operate around 350~MHz, but can otherwise be done with the system as-is.

\begin{figure*}
    \centering
    \includegraphics[width=1\columnwidth]{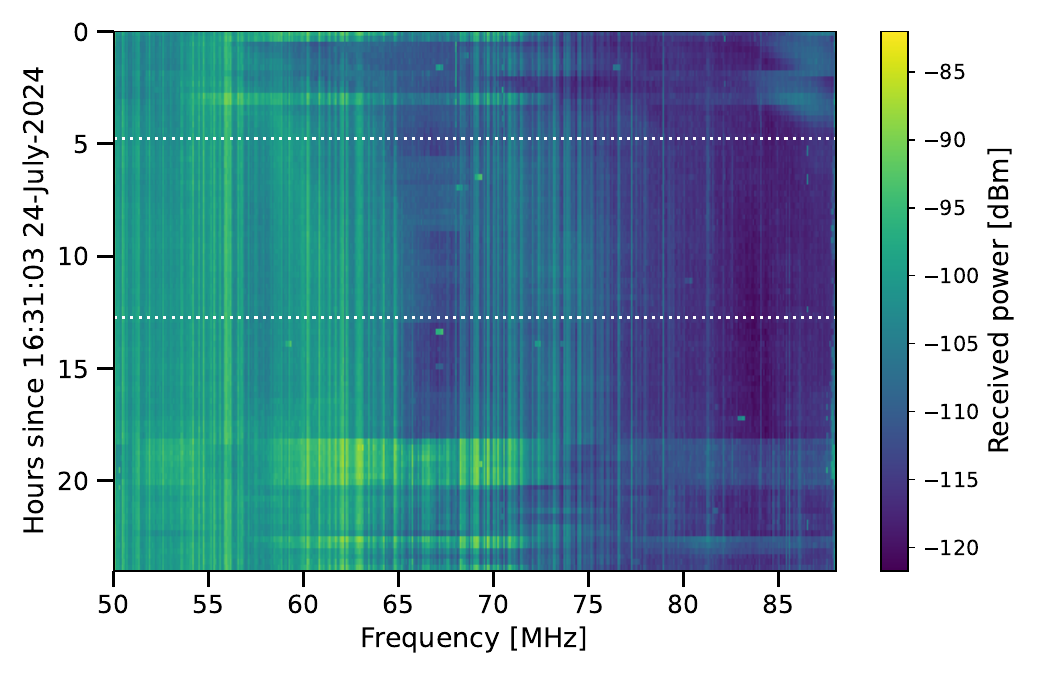}
    \includegraphics[width=1\columnwidth]{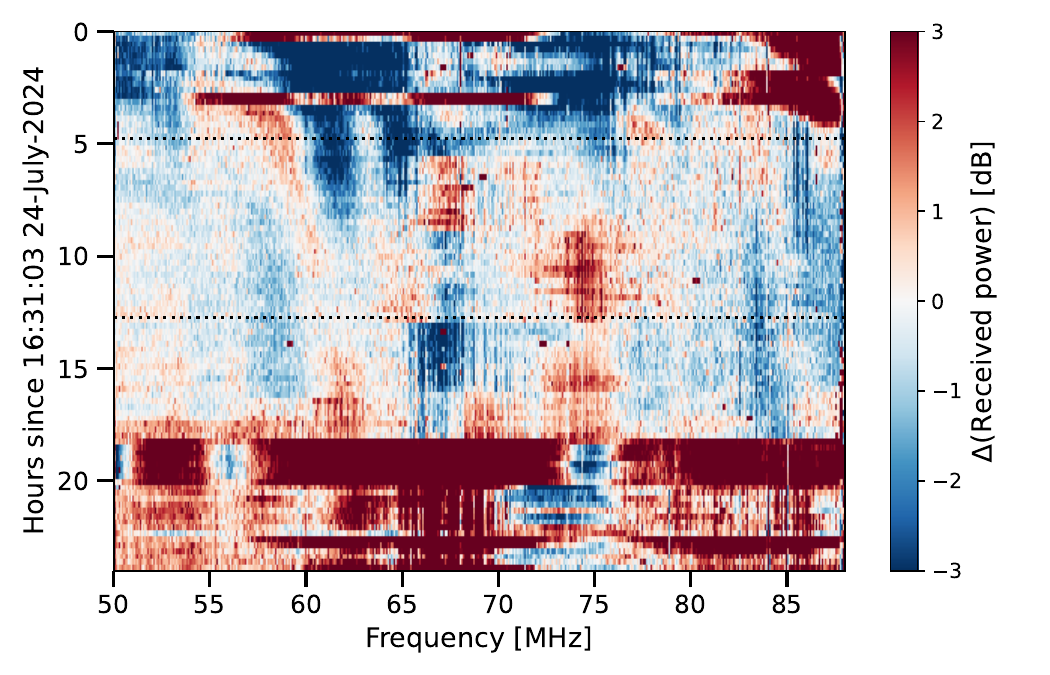}
    \caption{Waterfall plots showing 24 hours of RFI monitoring at the Jodrell Bank Observatory in the 55 -- 88 MHz band (left), and the same data with the median in each frequency channel subtracted (right). The colour scale of the latter is saturated to highlight deviations. The data were recorded using a commercial monitor with a wideband antenna at 15 min intervals from 16:31 (UTC+1) on 2024-07-24. The dotted lines show the local times of sunset and sunrise. (Data provided by S. Smith.)}
    \label{fig:rfi}
\end{figure*}

\paragraph*{Prototype horn:} The prototype antenna is a scaled-down pyramidal horn with a wooden frame and welded mesh conductive surface consisting of multiple overlapping segments. It is attached to a rectangular waveguide of similar construction, and targets a centre frequency of $\sim 350$~MHz. The right panel of Fig.~\ref{fig:prototype} shows the flare section during construction at JBO. The frame is made in several trapezoidal sections with reinforcing ribs inside, and joined together with metal plates and brackets. Inside the frame are multiple sections of 13mm galvanised welded mesh, stapled to the wooden frame, with several cm overlap between each section. The structure is positioned to point vertically at the zenith, with metal scaffolding poles used to support it. At the aperture, the conducting surface has a square profile of side length 2.19m, giving $\lambda / D \approx 22^\circ$ at 350~MHz. The purpose of the scaled-down prototype is to experiment with fabrication and beam measurement methods, to study the impact of environmental conditions (e.g. thermal expansion, wind loading, and soil moisture) on the real-world antenna pattern \citep[c.f.][]{2025MNRAS.538.1301P}.

\paragraph*{Site characteristics:} The JBO site is located approximately 27km south of Manchester, UK, a major population centre, and is not in a radio-quiet zone. As such, the local RFI environment can be expected to be challenging. We carried out 24h of monitoring with a 15 minute cadence, starting at 16:31 local time on 2024-07-24 using a commercial wideband monitor. Fig.~\ref{fig:rfi} shows the raw waterfall plot on the left, and the same data with the median in time subtracted from each frequency channel (as a rough bandpass estimate) on the right. The start of the FM band can be seen just at the top of the frequency range, and is well-localised in frequency. The narrow vertical stripes are well-removed by the median subtraction. While there remain a number of artefacts with different levels of localisation in time and frequency, the local night-time and early morning are relatively quiet. Further, deeper observations will establish whether the site is viable for global signal observations without needing to flag an excessive fraction of the data.

\begin{figure}
    \centering
    \includegraphics[width=1\columnwidth]{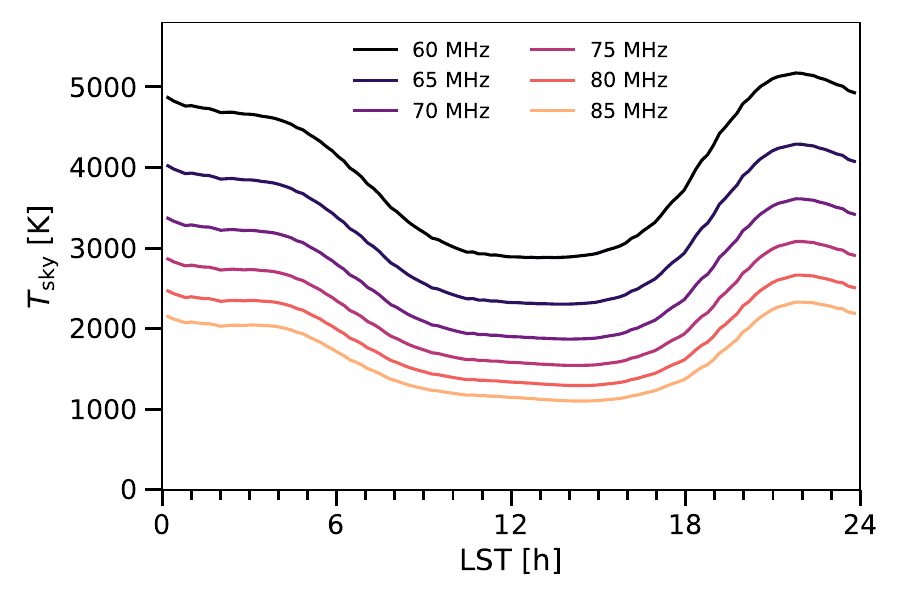}
    \caption{Sky temperature as a function of LST and frequency, estimated from the GSM 2016 sky model smoothed by a Gaussian beam with FWHM $\approx \lambda/D$ for $D = 7$m. The assumed site is Jodrell Bank Observatory.}
    \label{fig:tsky}
\end{figure}

\paragraph*{Observing field:} Being relatively far north ($53^\circ$N), the zenith pointing passes through a high Galactic latitude region with low synchrotron emission. Fig.~\ref{fig:tsky} shows the sky temperature at zenith as a function of frequency and LST from the JBO site, assuming a Gaussian beam with FWHM$\approx \lambda / D$ (with $D = 7$m). The GSM 2016 model has been used for this calculation, and should be treated with care as there are potentially significant uncertainties in the monopole temperature vs frequency. There is a $\sim\!14$-hour region in LST (from $\sim\! 6 - 20$~h) where the integrated sky temperature is around 2000~K below its peak in the centre of the design band. We plan to only observe at night; LSTs of 6~h and 13~h convert to about 23~h and 6~h GMT (UTC+0), resulting in a lengthy overnight window in which the sky temperature is lower, significantly reducing the thermal noise during this period.

\section{Conclusions} \label{sec:conclusions}

The redshifted 21cm global signal from neutral hydrogen is a particularly sought-after probe of the early Universe. As well as telling us about the thermal state of the intergalactic medium, and the timing and intensity of early heating and cooling processes that are central to theories of star and galaxy formation, it also has the potential to identify exotic energetic processes that may have happened at early times. The signal itself is minuscule however, of order tens to hundreds of millikelvin, as well as being spectrally relatively smooth in most models. This makes it extremely challenging to separate from vastly brighter (and comparably smooth) Galactic and extragalactic foregrounds. This is made all the more difficult by the complex, frequency-dependent spectral responses of radio telescopes, which modulate the foregrounds and result in spurious signals unless the instrument is calibrated and characterised with exquisite precision.

Painstaking efforts to develop sufficiently precise and characterisable hardware and data analysis techniques to overcome these issues appeared to come to fruition with the reported detection of an absorption feature around $z \approx 17$ by the EDGES experiment in Western Australia \citep{2018Natur.555...67B}. This feature was anomalously deep, however, and outside of the bounds of what could be realised within standard physical models of IGM physics. Either exotic cooling processes \citep[e.g.][]{Barkana2018} or a large background of radio photons \citep[e.g.][]{Feng2018} -- with caveats due to soft photon heating effects \citep{Acharya2023, Cyr2024} -- would need to be present at early times to explain the signal as observed. The latter may be directly related to the ARCADE-2 excess \citep{Fixsen2011, Dowell2018}, which is still lacking an explanation \citep[see][for broad discussion]{Singal2018, Singal2023}.
Alternatively, subtle problems with the data analysis or instrumental characterisation, coupled with the bright foregrounds, could have resulted in a spurious signal \citep[e.g.][]{2018Natur.564E..32H, 2018Natur.564E..35B, 2019ApJ...874..153B, 2020MNRAS.492...22S}. Observations in neighbouring frequency bands with EDGES appear to be consistent with the original detection claim \citep{2017ApJ...847...64M, 2018ApJ...863...11M}, while independent follow-up by the SARAS-3 instrument has not corroborated the EDGES signal \citep{2022NatAs...6..607S, 2022NatAs...6.1473B}. A number of other experiments are now under way to try to make sense of the situation, either by successfully re-observing the same feature as EDGES, or making their own independent detection of the 21cm global signal, whatever it may look like.

The task remains difficult, however. While the newer experiments can benefit from lessons learned through the operation of EDGES and SARAS, for instance, other difficulties remain, and may not yet be solvable at the level required to make a robust detection of the global signal. Examples include the difficulty of measuring the antenna pattern with high accuracy while it is deployed in the field, environmental factors that are highly complicated to measure and model, and various instrumental systematic effects related to the stability and spectral response of the receiver system. Owing to the use of instrumental designs and calibration methods that are similar in some important respects, independent experiments could still suffer from shared systematic effects, complicating attempts to decisively follow-up the EDGES detection.

With this in mind, in this paper we have put forward a proposal for an alternative type of 21cm global signal experiment that tries to use independent approaches from existing experiments wherever possible. Most notably, this involves using a large horn antenna instead of a more conventional compact design, such as a blade dipole. Horn antennas are a mature and well-understood technology that can be simulated with high accuracy. They are the go-to design for precision-calibrated applications at higher frequencies \citep[e.g.][]{1990ApJ...360..685S, 2005RScI...76l4703S, 2009JInst...4T2004V}, where they are smaller and thus easier to manufacture. By using a type of antenna that is more readily characterisable, and which should also be less sensitive to its immediate environment, we therefore hope to achieve better control over the antenna pattern than alternatives. The downside is the additional cost and complexity of having to build a large structure. In this paper, we presented a practical design and EM simulations for a large horn antenna that comes close to meeting our design requirements in the sub-100~MHz band. We also pointed out some deficiencies in the design however, that can be addressed through elaborations such as corrugations, aperture chokes, and similar.

As well as using a horn antenna, we have also proposed a novel continuous wave calibration strategy, intended to allow correlated gain fluctuations ($1/f$ noise) to be calibrated out without needing rapid switching between different signal paths, as in Dicke switching and related strategies. The practical effectiveness of this approach remains to be demonstrated for a relatively broadband receiver system like the one we propose, but has several potential advantages in terms of stability and efficiency.

We also used simple simulations to study the impact of using a narrower bandwidth receiver that is constrained by the need to filter the FM band at 87.5~MHz and above. While a broader bandwidth is always preferable, we found that the main hurdle is likely to be disentangling the foregrounds from the global signal regardless of bandwidth, as both signals are spectrally smooth and therefore have significant overlap in terms of their spectral behaviour. While a well-characterised horn antenna pattern may permit more accurate forward modelling of the foreground emission, there are still many significant sources of uncertainty, including the spatial and spectral behaviour of the foregrounds themselves. Various Bayesian approaches have been proposed to overcome this issue \citep[e.g.][]{2021MNRAS.506.2041A, 2022MNRAS.517.2264M, 2023MNRAS.521.3273S}.

As more `anomalies' continue to {\edit crop up in} observations of the frequency spectrum of the sub-GHz sky, it is becoming increasingly pressing to develop new precision-calibrated experiments to follow them up and further investigate possible causes. The RHINO concept presented here provides one practical approach to achieving the necessary calibration accuracy and dynamic range to study the EDGES signal and ARCADE-2 excess, and will provide a valuable cross-check on similar observations carried out via different observational approaches.

\balance

\section*{Acknowledgements}

We are grateful to many colleagues for valuable advice, discussions, and assistance, including J.~Barry, E.~Blackhurst, D.~Brown, I.~Browne, P.~Clarke, E.~de Lera Acedo, C.~Dickinson, A.~Galtress, S.~Garrington, J.~Hickish, A.~Holloway, J.~Kitching, N.~Mahesh, G.~Pisano, {\color{blue} D.~Price}, N.~Razavi-Ghods, S.~Smith, R.~Spencer, P.~Wilkinson, and M.~Wirawan.
This result is part of a project that has received funding from the European Research Council (ERC) under the European Union's Horizon 2020 research and innovation programme (Grant agreement No. 948764; PB). We acknowledge support from STFC grant ST/Y509942/1.
This work used the DiRAC@Durham facility managed by the Institute for Computational Cosmology on behalf of the STFC DiRAC HPC Facility (www.dirac.ac.uk). The equipment was funded by BEIS capital funding via STFC capital grants ST/P002293/1, ST/R002371/1 and ST/S002502/1, Durham University and STFC operations grant ST/R000832/1. DiRAC is part of the National e-Infrastructure.

We acknowledge use of the following software: 
{\tt matplotlib} \citep{matplotlib}, {\tt numpy} \citep{numpy}, and {\tt scipy} \citep{2020SciPy-NMeth}.

\section*{Data Availability}

Data used in this paper, including the simulated antenna patterns, are available from \url{https://github.com/RHINO-Experiment}. Current versions of the RHINO simulation, calibration, and control software are also available from this repository, where they are in active development.



\bibliographystyle{mnras}
\bibliography{rhino} 




\label{lastpage}
\end{document}